\documentclass[aps,prb,showkeys,a4paper,floatfix,amsmath,amssymb,superscriptaddress]{revtex4}
\usepackage{natbib}
\usepackage{dcolumn}
\usepackage{graphicx}
\usepackage{epsfig}

\begin{document}
\title{Simulation of Quantum Computation: A deterministic event-based approach%
\footnote{To appear in: Journal of Computational and Theoretical Nanoscience}%
}

\author{K. Michielsen}
\email{kristel@phys.rug.nl}
\affiliation{Department of Applied Physics,
Materials Science Centre, University of Groningen, Nijenborgh 4,
NL-9747 AG Groningen, The Netherlands
}
\author{K. De Raedt}
\email{deraedt@cs.rug.nl}
\affiliation{Department of Computer Science,
University of Groningen, Blauwborgje 3,
NL-9747 AC Groningen, The Netherlands}
\author{H. De Raedt}
\email{deraedt@phys.rug.nl}
\homepage{http://www.compphys.org}
\affiliation{Department of Applied Physics,
Materials Science Centre, University of Groningen, Nijenborgh 4,
NL-9747 AG Groningen, The Netherlands
}
\begin{abstract}
We demonstrate that locally connected networks of machines
that have primitive learning capabilities
can be used to perform a deterministic, event-based simulation
of quantum computation.
We present simulation results for basic quantum operations such
as the Hadamard and the controlled-NOT gate, and
for seven-qubit quantum networks that implement
Shor's numbering factoring algorithm.
\end{abstract}
\date{\today}

\maketitle

\def\ORDER#1{\hbox{${\cal O}(#1)$}}
\def\BRA#1{\langle #1 \vert}
\def\KET#1{\vert #1 \rangle}
\def\EXPECT#1{\langle #1 \rangle}
\def\BRACKET#1#2{\langle #1 \vert #2 \rangle}
\def\hbar{{\mathchar'26\mskip-9muh}}
\def\mod{{\mathop{\hbox{mod}}}}
\def\CNOT{{\mathop{\hbox{CNOT}}}}
\def\Tr{{\mathop{\hbox{Tr}}}}
\def\bPsi{{\mathbf{\Psi}}}
\def\bPhi{{\mathbf{\Phi}}}
\def\bzero{{\mathbf{0}}}
\def\Eq#1{(\ref{#1})}
\def\NOBAR#1{#1}
\def\BAR#1{\overline{#1}}

\def\DLM{DLM}
\def\DLMS{DLMs}

\section{Introduction}\label{intro}

Recent advances in nanotechnology are paving the way to attain
control over individual microscopic objects~\cite{CHIO04,RUGA04,ELZE04,XIAO04,FUJI03}.
The ability to prepare, manipulate, couple and measure single quantum systems is essential
for quantum computation~\cite{NIEL00}.
Candidate systems to implement a quantum computer are ions or atoms in electromagnetic or optical traps,
photons in cavities, nuclear spins on molecules, quantum dots and superconductors.
These technological developments facilitate the study of single quantum systems
at the level of individual events.
Such experiments address the most fundamental aspects of quantum theory.
Indeed, quantum theory gives us only a recipe to compute the frequencies for
observing events. It does not describe individual events, such as
the arrival of a single electron at a particular position 
on the detection screen~\cite{FEYN65,HOME97,TONO98,BALL03}.
Reconciling the mathematical formalism (that does not describe single events) with
the experimental fact that each observation yields a definite
outcome is often referred to as the quantum measurement paradox.
This is the fundamental problem in the foundation
of quantum theory~\cite{FEYN65,HOME97,PENR90}.

In view of this, it is not a surprise that some of
the most fundamental experiments in quantum physics
have not been simulated in the event-by-event manner in which the experimental observations
are actually recorded~\cite{PerfectExperiments}.
One of the examples are the two-slit experiments in which the
individual electron counts build up the interference pattern~\cite{TONO98}.
Other examples are the single-photon
beam splitter and Mach-Zehnder interferometer experiments~\cite{GRAN86}.
Within the standard formalism of quantum theory, no algorithm has been found to perform
an event-based simulation of definite individual outcomes in quantum experiments~\cite {HOME97}.

In this paper, we take the point of view that a physical theory such as
quantum theory, is a specification of an algorithm to compute numbers
that can be compared to experimental data~\cite{KAMP88}.
Thinking in terms of algorithms opens new possibilities
to simulate phenomena for which a proper physical theory is not (yet) available.
As discussed before, quantum theory is unable to describe individual events in an
experiment but it provides an algorithm to compute the final, collective outcome.
We have already demonstrated that it is possible to construct deterministic processes
that generate events at a rate that
agrees with the quantum mechanical probability distribution,
without using quantum theory~\cite{KOEN04}.
In this paper we apply these concepts to quantum computation.
We present results of event-based simulations of
single-qubit quantum interference (Mach-Zehnder interferometer),
a two-qubit quantum circuit (controlled-NOT gate), and Shor's algorithm~\cite{SHOR99} to factorize $N=15$
on a seven qubit quantum computer.

The event-based simulation method that we describe
in this paper is not a proposal for another interpretation of quantum mechanics.
To avoid misunderstandings, we emphasize that our approach is not an extension
of quantum theory.
We simulate quantum systems without making use of the rules (algorithms) quantum theory
provides us. However, the final, collective results of our event-by-event deterministic, causal
learning processes are in perfect agreement with the probability distributions of quantum theory~\cite{PENR90}.
The event-based simulations build up the final outcome event-by-event,
just like in real experiments.
In physics terminology, the entire approach is particle-like
and satisfies Einstein's criteria of realism and causality~\cite{HOME97}.
In this sense, our approach can be viewed as a recipe
to construct event-based systems, that is ``classical'' models, that
behave as if they were ``quantum mechanical''.

The paper is organized as follows.
In Section~\ref{sec2} we briefly recall some basic elements of quantum computation.
We describe the Hadamard, controlled-NOT (CNOT) and Toffoli gate,
involving one, two or three qubits, repectively.
We discuss the quantum network for the Mach-Zehnder interferometer and for
Shor's quantum algorithm to factorize $N=15$ on a seven-qubit quantum computer.

In Section~\ref{sec3} we introduce the deterministic learning machine (DLM)
that is at the core of the event-based simulation approach.
We present a mathematical analysis that proves that these machines
generate events at a rate that agrees with the corresponding
quantum mechanical probabilities.
One of the essential features of (networks of) \DLMS\ is that they process one event at a time.
After applying a deterministic decision process to the input event and sending out an output
event, a new input event can be processed. Another essential feature of \DLMS\ is that they
do not store information about individual events.
Networks of \DLMS\ are capable of unsupervised learning~\cite{KOEN04}
but they have very little in common with neural networks~\cite{HAYK99}.
The sequence of events that is generated by a DLM is stricly deterministic. This is modified in
the stochastic learning machine (SLM). The event-by-event learning processes in a SLM are still
deterministic and causal but the output events are randomly distributed.
This modification is necessary if we want to mimic
the apparent random order in which quantum
events are detected in experiments~\cite{GRAN86,TONO98}.

In Section~\ref{sec4} we first describe the construction of DLM networks and
present results from the event-based simulation of the Hadamard gate and the Mach-Zehnder interferometer,
the latter showing that DLM-based networks correctly reproduce quantum interference phenomena.
Then we describe how to simulate a CNOT gate using DLM and SLM networks.
Finally, we present the simulation results of Shor's number factoring algorithm~\cite{SHOR99} implemented
on DLM and SLM networks.
A summary and outlook is given in Section~\ref{SUMM}.

\section{Quantum Computation}\label{sec2}

This section summarizes those aspects of quantum computation that
are necessary to understand the examples of quantum algorithms
that we use in Section~\ref{sec4} to demonstrate
that local, causal and deterministic processes
can simulate quantum computers on an event-by-event basis.

\subsection{Preliminaries}

The state of an elementary storage unit of a quantum computer, the quantum bit or qubit,
is described by a two-dimensional vector of Euclidean length one.
Denoting two orthogonal basis vectors of the
two-dimensional vector space by $\KET{0}$ and $\KET{1}$, the state $\KET{\Phi}$
of the qubit can be written as a linear superposition of the basis states
$\KET{0}$ and $\KET{1}$:
\begin{equation}
\KET{\Phi}=a_0\KET{0} +a_1\KET{1},
\label{STAT0}
\end{equation}
where $a_0$ and $a_1$ are complex numbers such that $|a_0|^2+|a_1|^2=1$.
The appearance of complex numbers suggests
that one qubit can contain an infinite amount of information.
However, it is impossible to retrieve all this information~\cite{SCHI68,BAYM74,BALL03}.
The result of inquiring about the state of the qubit, that is the outcome of a measurement,
is either 0 or 1.
The frequency of obtaining 0 (1) can be estimated
by repeated measurement of the same state of the
qubits and is given by $|a_0|^2$ ($|a_1|^2$)~\cite{SCHI68,BAYM74,BALL03}.

According to quantum theory~\cite{QuantumTheory}, the internal state of a quantum computer
with $L$ qubits is described by a unit vector (state vector) in a $D=2^{L}$
dimensional space (of complex numbers)~\cite{NIEL00}.
Adopting the convention of quantum computation literature~\cite{NIEL00},
the state of an $L$-qubit quantum computer is represented by
\begin{eqnarray}
\KET{\Phi}&=&a({0\ldots00}) \KET{0\ldots00}
+a({0\ldots01}) \KET{0\ldots01}
+\ldots
+a({1\ldots10}) \KET{1\ldots10}
+a({1\ldots11}) \KET{1\ldots11}
\nonumber \\
&=&a_0 \KET{0}
+a_1 \KET{1}
+\ldots
+a_{2^L-2} \KET{2^L-2}
+a_{2^L-1} \KET{2^L-1}
,
\label{STAT2}
\end{eqnarray}
where in the last line of Eq.~\Eq{STAT2}, the binary representation of the integers
$0,\ldots,2^{L-1}$ was used to denote
$\KET{0}\equiv\KET{0\ldots00},\ldots,\KET{2^L-1}\equiv\KET{1\ldots11}$ and
$a_0\equiv a({0\ldots00}),\ldots,a_{2^L-1}\equiv a({1\ldots11})$.
As usual, we normalize the state vector, that is $\BRACKET{\Phi}{\Phi}=1$,
by rescaling the complex-valued amplitudes $a_{i}$ according to
\begin{eqnarray}
\sum_{i=0}^{2^L-1}|a_i|^2=1
.
\label{STAT1}
\end{eqnarray}

The internal state of the quantum computer evolves
in time according to a sequence of unitary transformations~\cite{NIEL00}.
A quantum algorithm is a sequence of such unitary operations.
Of course, not every sequence corresponds to a meaningful computation.
Furthermore, if a quantum algorithm cannot exploit the fact that
the intermediate state of the quantum computer is described by a linear
superposition of basis vectors, it will not be faster than its classical
counterpart.
As the unitary transformation may change all amplitudes simultaneously,
a quantum computer is a massively parallel machine~\cite{NIEL00}, at least
in theory.

It has been shown that an arbitrary unitary operation
can be written as a sequence of single qubit operations
and the controlled-NOT (CNOT) operation on two qubits~\cite{DIVI95a,NIEL00}.
Therefore, single-qubit operations and the CNOT operation
are sufficient to construct a universal quantum computer~\cite{NIEL00}.
The final state of the quantum computer can be calculated by performing
the (unitary) matrix-vector multiplications
that correspond to the application of this sequence (determined by
the quantum algorithm) of single-qubit and CNOT operations.
According to quantum theory, the squares of the absolute values of the
elements of the state vector are the probabilities for observing
the quantum computer in one of its $2^{L}$ states.

The unitary time evolution of the internal state of the quantum computer
is interrupted at the point where we inquire about the value
of the qubits, that is as soon as we perform a measurement on the qubits.
If we perform the readout operation on a qubit, we get a definite answer, either 0 or 1,
and the information encoded in the superposition is lost.
The process of measurement cannot be described by a unitary transformation~\cite{HOME97,BALL03}.
Therefore, we do not consider it to be part of a quantum algorithm.

\subsection{Single-qubit operations}\label{CIRC}\label{ILLU}\label{QB}

\setlength{\unitlength}{1cm}
\begin{figure*}[t]
\begin{center}
\includegraphics[width=12cm]{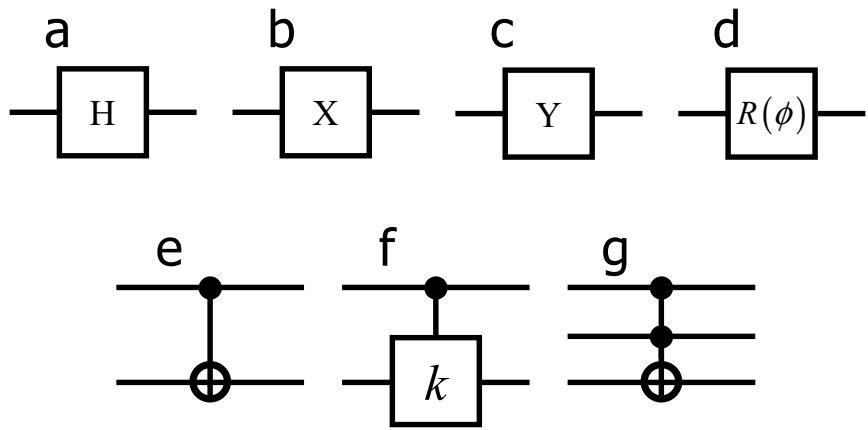}
\end{center}
\caption{
Graphical representation of some of the basic gates used in quantum computation;
(a) Hadamard gate;
(b) Rotation by $\pi/2$ about the $x$-axis;
(c) Rotation by $\pi/2$ about the $y$-axis;
(d) Single qubit phase shift by $\phi$;
(e) CNOT gate;
(f) Controlled phase shift by $\phi =\pi /k$;
(g) Toffoli gate.
The horizontal lines denote the qubits involved in the quantum operations.
The dots and crosses denote the control and target qubits, respectively.}
\label{fig:Gates}
\end{figure*}

In general, a qubit can be represented by a spin-1/2 system.
The state $\KET{\Phi}$ of a qubit (see Eq.(\ref{STAT0})) can
therefore also be written as a linear combination of
the spin-up and spin-down states~\cite{SCHI68,BAYM74,BALL03}:
\begin{equation}
\KET{\Phi}=a_0\KET{\uparrow}+a_1\KET{\downarrow}
,
\label{SPIN1}
\end{equation}
where~\cite{NIEL00}
\begin{equation}
\KET{0}=\KET{\uparrow}=
\left(
\begin{array}{c}
1 \\
0
\end{array}
\right)
\quad,\quad \KET{1}=\KET{\downarrow}=
\left(
\begin{array}{c}
0 \\
1
\end{array}
\right)
.
\label{SPIN4}
\end{equation}
The three components of the spin-1/2 operator ${\bf S}=(S^x,S^y,S^z)$
are defined (in units such that $\hbar=1$) by~\cite{SCHI68,BAYM74,BALL03}
\begin{equation}
S^x=\frac{1}{2}
\left(
\begin{array}{cc}
0&1 \\
1&0
\end{array}
\right),\quad
S^y=\frac{1}{2}
\left(
\begin{array}{cc}
\phantom{-}0&-i \\
\phantom{-}i&\phantom{-}0
\end{array}
\right),\quad
S^z=\frac{1}{2}
\left(
\begin{array}{cc}
\phantom{-}1&\phantom{-}0 \\
\phantom{-}0&-1
\end{array}
\right)
,
\label{SPIN2}
\end{equation}
and are chosen such that
$\KET{\uparrow}$ and $\KET{\downarrow}$ are eigenstates of $S^z$ with
eigenvalues +1/2 and -1/2, respectively.

The expectation values of the three components of the qubits are
defined as
\begin{equation}
\EXPECT{Q^\alpha}=1/2-\EXPECT{S^\alpha}\quad,\quad \alpha=x,y,z,
\label{SPIN6}
\end{equation}
where $\EXPECT{A}=\BRACKET{\Phi}{A|\Phi}/\BRACKET{\Phi}{\Phi}$.
A qubit is in the state $\KET{0}$ or $\KET{1}$ if $\EXPECT{Q^z}=0$ or $\EXPECT{Q^z}=1$, respectively.

A rotation of the state Eq.\Eq{SPIN1} about a vector ${\bf v}$ corresponds
to the unitary matrix
\begin{equation}
e^{i{\bf v}\cdot{\bf S}}=\openone\cos\frac{v}{2}+\frac{2i{\bf v}\cdot{\bf S}}{v}\sin\frac{v}{2}
,
\label{SPIN11}
\end{equation}
where $\openone$ denotes the unit matrix and $v=\sqrt{v_x^2+v_y^2+v_z^2}$ is the length of the
vector ${\bf v}$.

For later reference, it is useful to list a few
special cases of Eq.\Eq{SPIN11}.
The Hadamard operation $H$ and rotations $X$ and $Y$ of the state vector
by $\pi/2$ about the $x$ and $y$-axis, respectively,
are defined by~\cite{NIEL00}
\begin{equation}
H\equiv\frac{1}{\sqrt{2}}
\left(
\begin{array}{cc}
\phantom{-}1&\phantom{-}1 \\
\phantom{-}1&-1
\end{array}
\right),\quad
X\equiv e^{i\pi S^x/2}=\frac{1}{\sqrt{2}}
\left(
\begin{array}{cc}
1&i \\
i&1
\end{array}
\right)
,\quad
Y\equiv e^{i\pi S^y/2}=\frac{1}{\sqrt{2}}
\left(
\begin{array}{cc}
\phantom{-}1&\phantom{-}1 \\ -1&\phantom{-}1\\
\end{array}
\right)
.
\label{SPIN10}
\end{equation}
We also introduce the symbol
\begin{equation}
R(\phi)=e^{i\phi/2}e^{-i\phi S^z}=
\left(
\begin{array}{cc}
\phantom{-} 1&\phantom{-} 0\\
\phantom{-} 0&\phantom{-} e^{i\phi}
\end{array}
\right)
,
\label{SING6}
\end{equation}
to represent the single-qubit phase-shift operation by a phase $\phi$.
The graphical symbols of $H$, $X$, $Y$, and $R(\phi)$ are shown in Fig.~\ref{fig:Gates}.
The inverse of a unitary operation $U$ is denoted by $\BAR U$.

\setlength{\unitlength}{1cm}
\begin{figure*}[t]
\begin{center}
\begin{picture}(14,7)
\put(-1.5,0){\includegraphics[width=8cm]{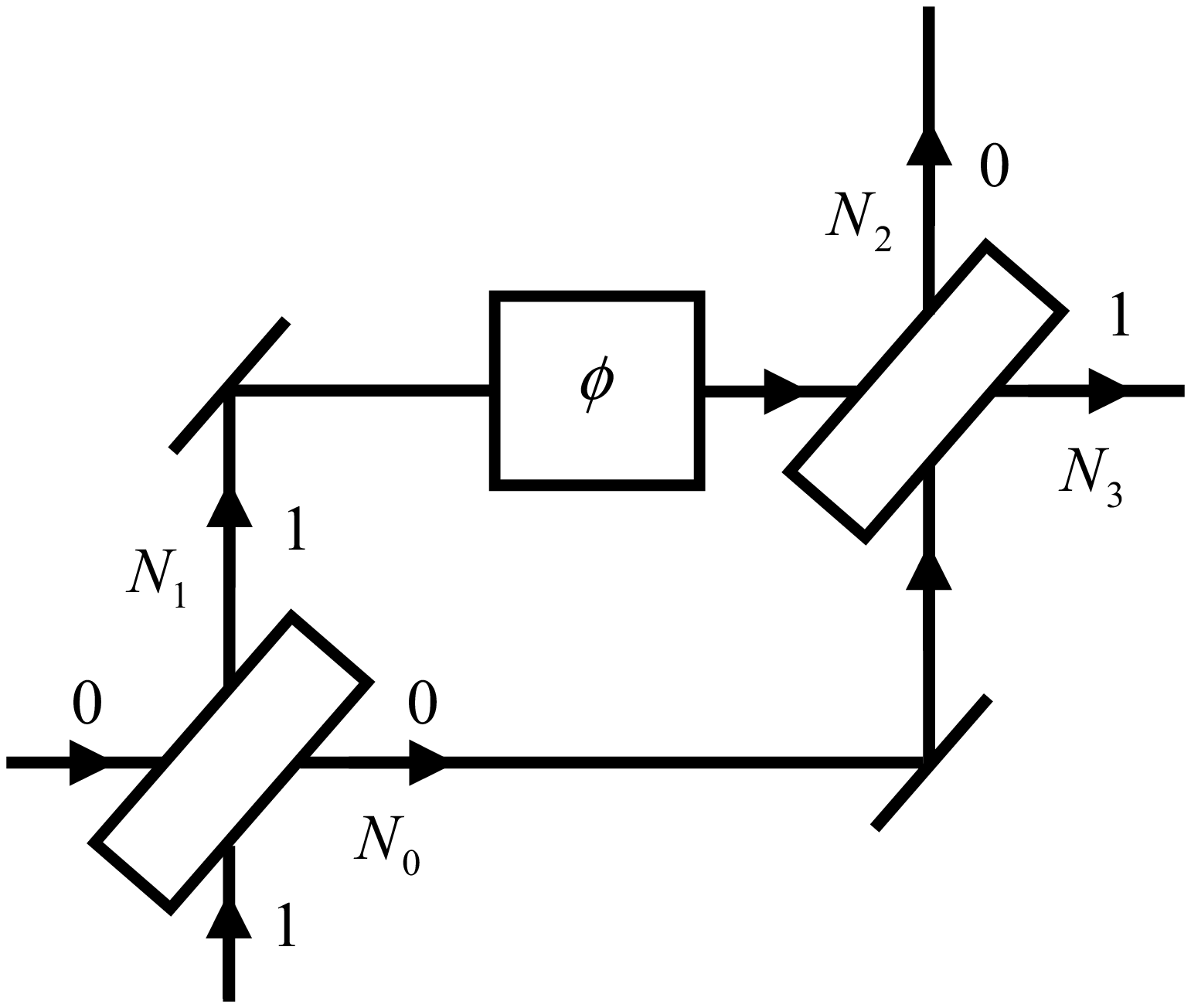}}
\put(7,2){\includegraphics[width=9cm]{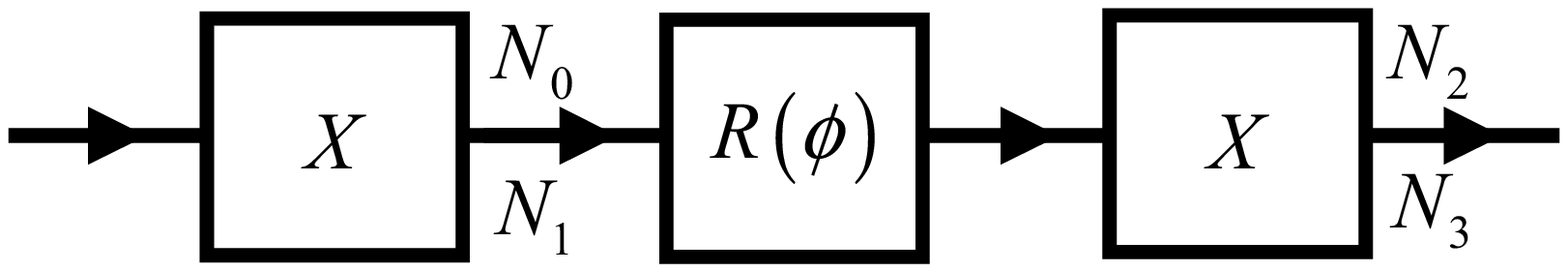}}
\end{picture}
\caption{
Left:
Diagram of the Mach-Zehnder interferometer,
consisting of two beam splitters, two mirrors,
and a device that changes the phase of the light wave by $\phi$.
Right:
Diagram of the equivalent quantum network.
}
\label{one-mzi}
\end{center}
\end{figure*}

The Mach-Zehnder interferometer is a simple but nontrivial
example of a single-qubit system in which the information
contained in the phase of the wave function is essential~\cite{BORN64,GRAN86,RARI97,NIEL00}.
The schematic layout of the apparatus is shown in Fig.~\ref{one-mzi}.
$N$ photons enter the first beam splitter through the input channels 0 or 1.
The beam splitter distributes the photons over its
two output channels (0 or 1):
The number of photons in these
channels is $N_0$ and $N_1$, respectively.
If there are no photons in one of the input channels,
the beam splitter equally divides the photons
over its output channels, that is $N_0=N_1=N/2$.
The photons then propagate and experience a phase shift of $\phi$
if they left the beam splitter via channel 1.
The photons are collected at another beam splitter.
Finally, detectors count the number of photons in the two output channels
0 and 1 of the second beam splitter.
The number of photons in these channels is denoted by
$N_2$ and $N_3$, respectively.
We assume that no photons are lost in this process so that $N=N_0+N_1=N_2+N_3$.

The relation between input and output amplitudes
of the Mach-Zehnder interferometer is given by~\cite{BORN64,GRAN86,RARI97,NIEL00}
\begin{eqnarray}
\left(
\begin{array}{c}
b_0\\
b_1
\end{array}
\right)
=
XR(\phi)X
\left(
\begin{array}{c}
a_0\\
a_1
\end{array}
\right)
=
\frac{1}{2}
\left(
\begin{array}{cc}
1&i\\
i&1
\end{array}
\right)
\left(
\begin{array}{cc}
1&0\\
0&e^{i\phi}
\end{array}
\right)
\left(
\begin{array}{cc}
1&i\\
i&1
\end{array}
\right)
\left(
\begin{array}{c}
a_0\\
a_1
\end{array}
\right)
,
\label{MZI2}
\end{eqnarray}
where ($a_0,a_1)$  and ($b_0,b_1)$ are the complex-valued amplitudes
at the input and output, respectively.
The $X$ operations represent the beam splitters while the phase shift
$R(\phi)$ mimics the effect of changing the optical path length.

Let us consider the case where all the photons
enter the interferometer through channel 0.
Assuming that the photons originate from
a coherent source, we have $(a_0,a_1)=(\cos\psi_0+i\sin\psi_0,0)$.
From Eq.\Eq{MZI2} it follows that
the probabilities for observing a photon at one of the
output channels is given by
\begin{eqnarray}
|b_0|^2=\sin^2\frac{\phi}{2},\quad
|b_1|^2=\cos^2\frac{\phi}{2}.
\label{MZ3}
\end{eqnarray}

\subsection{Two qubits: CNOT operation and controlled phase shift}\label{CNOT}

\begin{figure*}[t]
\begin{center}
\includegraphics[width=14cm]{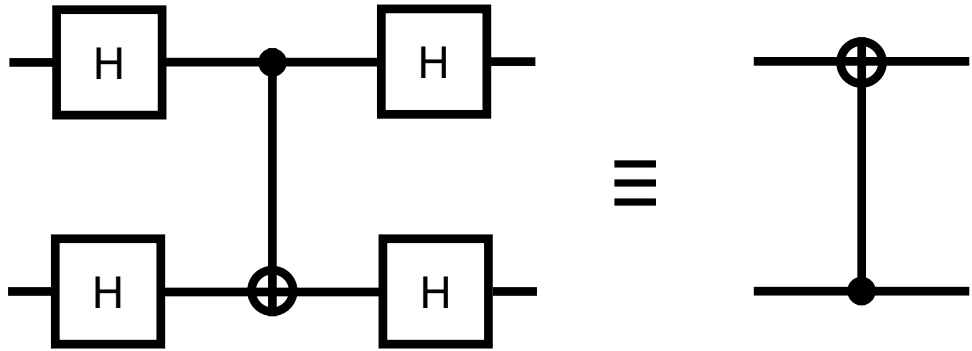}
\caption{
Quantum circuit representation of a CNOT
gate where the second (bottom)
qubit acts as control qubit and the first
(top) qubit is the target qubit.
}
\label{figcnot2}
\end{center}
\end{figure*}

Computation requires some form of communication between the qubits.
It has been shown that any form of communication
between qubits can be reduced to a combination of single-qubit operations
and the CNOT operation on two qubits~\cite{DIVI95a,NIEL00}.
By defintion, the CNOT gate flips the target qubit if
the control qubit is in the state $\KET{1}$~\cite{NIEL00}.
If we take the first qubit
(that is the least significant bit in the binary notation of an integer)
as the control qubit, we have
\begin{eqnarray}
\CNOT_{21}\KET{\Phi}&=&\CNOT_{21}
(a_0\KET{00}+a_1\KET{01}+a_2\KET{10}+a_3\KET{11})
\nonumber\\
&=&a_0\KET{00}+a_3\KET{01}+a_2\KET{10}+a_1\KET{11}
\nonumber\\
&=&a_0\KET{0}+a_3\KET{1}+a_2\KET{2}+a_1\KET{3}
\nonumber\\
&=&a_0\KET{0}_1\KET{0}_2+a_3\KET{1}_1\KET{0}_2
  +a_2\KET{0}_1\KET{1}_2+a_1\KET{1}_1\KET{1}_2.
\label{CNOT2}
\end{eqnarray}
where $a_0,\ldots ,a_3$ are the probability amplitudes
of the four different states and
$\KET{0}_i$ and $\KET{1}_i$ represent the
$\KET{0}$ and $\KET{1}$ state of the $i$-th qubit, respectively.
The graphical symbol of the CNOT operation is
shown in Fig.~\ref{fig:Gates}e.
The dot (cross) denotes the control (target) qubit.
A CNOT gate in which the control and target qubit are interchanged
can be built from four Hadamard gates and one CNOT gate~\cite{NIEL00}.
The quantum circuit for this ``reversed'' CNOT gate is shown in Fig.~\ref{figcnot2}.

The CNOT operation is a special case of the controlled
phase shift operation $R_{ji}(\phi)$.
The controlled phase shift operation with control qubit 1
and target qubit 2 reads

\begin{eqnarray}
R_{21}(\phi)\KET{\Phi}&=&
R_{21}(\phi)(a_0\KET{00}+a_1\KET{01}+a_2\KET{10}+a_3\KET{11}),
\nonumber\\
&=&a_0\KET{00}+\frac{(1+e^{i\phi})a_1+(1-e^{i\phi})a_3}{2}\KET{01}
+a_2\KET{10}+
\frac{(1-e^{i\phi})a_1+(1+e^{i\phi})a_3}{2}\KET{11}
\label{COMM4}
.
\end{eqnarray}
Graphically, the controlled phase shift $R_{ji}(\phi=\pi/k)$
is represented by a vertical line connecting a dot (control bit)
and a box denoting a single qubit phase shift by $\pi /k$ (see Fig.~\ref{fig:Gates}f).

\setlength{\unitlength}{1cm}

\subsection{Three qubits: Toffoli gate}\label{TOFF}

The Toffoli gate is a generalization of the CNOT gate
in the sense that it has two control qubits and one target qubit~\cite{BARE95,NIEL00}.
The target qubit flips if and only if the two control qubits are set.
Symbolically the Toffoli gate is represented by a vertical line
connecting two dots (control bits) and
one cross (target bit), as shown in Fig.~\ref{fig:Gates}g.

\begin{figure*}[t]
\begin{center}
\includegraphics[width=16cm]{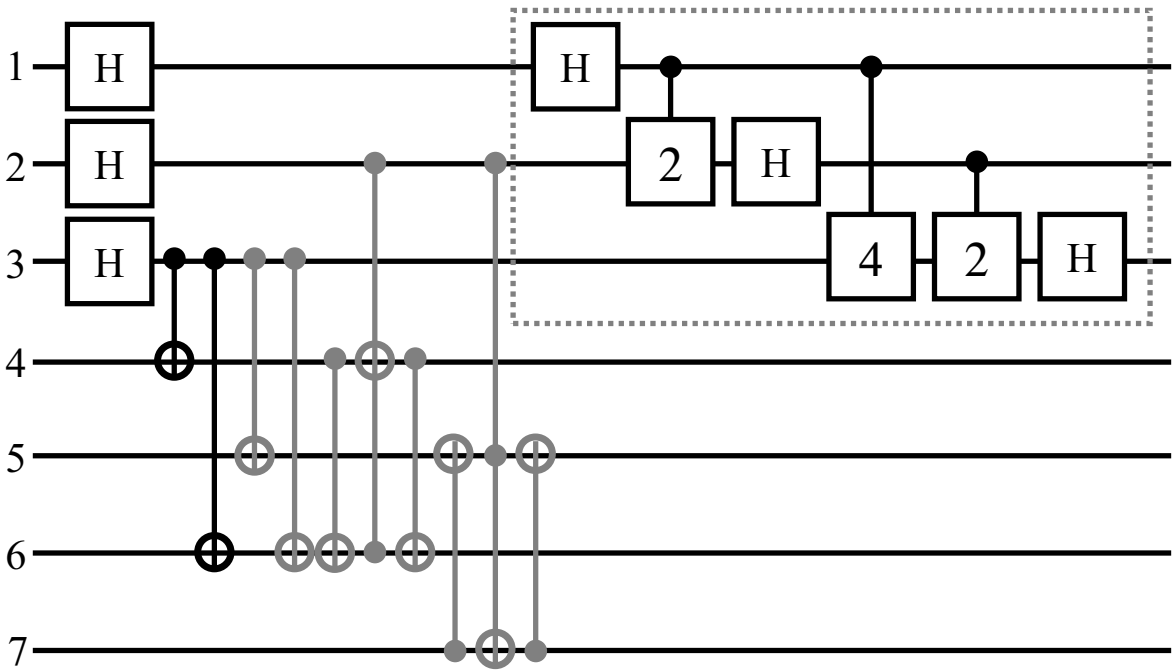}
\end{center}
\caption{Quantum network for Shor's quantum algorithm to find the factors of the number $N=15$
for the case $a=11$ (black colored CNOT gates only) and $a=7$
(gray colored CNOT and Toffoli gates only)~\cite{SYPE01b}.
$H$ denotes the Walsh-Hadamard transform.
The operations in the dashed box perform a 3-qubit Fourier transform~\cite{NIEL00}.
The operations ``2'' and ``4'' perform controlled phase shifts
with angles $\pi/2$ and $\pi/4$, respectively.
The other gates perform two-qubit (CNOT) or three-qubit (Toffoli) operations.
The initial state of the seven qubits of the quantum computer is (0,0,0,0,0,0,1).
}
\label{shor-circuit}
\end{figure*}

\subsection{Seven qubits: Factoring $N=15$ using Shor's algorithm}\label{SHOR}

We now consider the problem of factoring integers.
For the case $N=15$, an experimental realization
of this quantum algorithm on a seven-qubit NMR quantum computer
is described in Refs.~\onlinecite{SYPE01b} and \onlinecite{SYPE01a}.
The theory behind Shor's algorithm has been discussed at
great length elsewhere~\cite{NIEL00,SHOR99,EKER96}.
Therefore, we only recall the basic elements of Shor's
algorithm and focus on the implementation of the algorithm for the case $N=15$.
The theory in this section closely follows Ref.~\onlinecite{SYPE01b}.

Shor's algorithm is based on the fact that the factors $p$ and $q$ of an integer
$N=pq$ can be deduced from the period $M$ of the function $f(j)=a^j \mod N$
for $j=0,\ldots,2^n-1$ where $N\le 2^n$.
Here $a<N$ is a random number that has no common factors with $N$.
Once $M$ has been determined, at least one factor of $N$ can be found by computing
the greatest common divisor (g.c.d.) of $N$ and $a^{M/2}\pm 1$.

For $N=15$, the calculation of the modular exponentiation $a^j \mod N$ is almost trivial.
Using the binary representation of $j$ we can write
$a^j \mod N= a^{2^{n-1}j_{n-1}}\ldots a^{2j_{1}} a^{j_{0}}\mod N
= (a^{2^{n-1}j_{n-1}}\mod N) \ldots (a^{2j_{1}}\mod N) (a^{j_{0}}\mod N) \mod N$,
showing that we only need to implement $(a^{2^{k}j_{k}}\mod N)$.
For $N=15$ the allowed values for $a$ are $a=2,4,7,8,11,13,14$.
If we pick $a=2,7,8,13$ then $a^{2^k} \mod N=1$ for all $k>1$.
For the remaining cases we have $a^{2^k} \mod N=1$ for all $k>0$.
Thus, for $N=15$ only two (not four) qubits are sufficient
to obtain the period of $f(j)=a^j \mod N$~\cite{SYPE01b}.
As a matter of fact, this analysis provides enough information
to deduce the factors of $N=15$ using Shor's procedure
so that no further computation is necessary.
Nontrivial quantum operations
are required if we decide to use three (or more) qubits to determine
the period of $f(j)=a^j \mod N$~\cite{SYPE01b}.
Following Ref.~\onlinecite{SYPE01b}, we will consider a seven-qubit quantum computer
 with four qubits to hold $f(j)$
and three qubits to perform the Fourier transform to determine the period $M$.

The quantum circuit that implements Shor's algorithm that factors $N=15$
using $a=7$ and $a=11$ is depicted in Fig.~\ref{shor-circuit}~\cite{SYPE01b}.
Qubits 1 to 3 and 4 to 7 are used as registers to represent $j$
and $f(j)=a^j \mod N$, respectively.
Here, qubits 3 and 7 are the least significant qubits of
these two registers.
The initial state of the qubits is $(0,0,0,0,0,0,1)$,
that is, all qubits are prepared in the state $\KET{0}$ except
for qubit 7 which is prepared in the state $\KET{1}$.

The quantum networks to compute $a^j \mod 15$ for $j=0,\ldots,7$ and a fixed
input $a$ are easy to construct.
Examples for $a=7$ (six CNOT and two Toffoli gates)
and $a=11$ (two CNOT gates) are included in Fig.~\ref{shor-circuit}~\cite{SYPE01b}.
For example, consider the case $a=11=\KET{1011}$.
If $j$ is odd then $11^j \mod 15=11$ and the network should leave $\KET{1011}$ unchanged.
Otherwise, $11^j \mod 15=1$ and hence it should return $\KET{0001}$.
The network for this operation consists of two CNOT gates that have
as control qubit, the same least-significant qubit
(qubit 3 in Fig.~\ref{shor-circuit}) of the three qubits
that are input to the Fourier transform.
The sequence of CNOT and Toffoli gates that performs similar operations
for the other cases can be found in the same manner.
In the NMR implementation, additional simplications of the $a=7$ circuit
where necessary to render the experiment feasible~\cite{SYPE01b}.
There is no need to do this here, so we use the circuits as shown in
Fig.~\ref{shor-circuit}.
Elsewhere, we describe a quantum computer emulator (QCE) that
simulates models of ideal and physically realizable quantum computers~\cite{RAED00,MICH03,RAED05}.
The software distribution of QCE~\cite{QCEDOWNLOAD}
contains an implemention of the circuits shown in Fig.~\ref{shor-circuit},
demonstrating that these circuits work correctly.

For the quantum network of Fig.~\ref{shor-circuit},
we can determine the period $M$ of the function $f(j)=a^j \mod N$
from the expectation values of the first three qubits.
The state of the quantum computer before it starts performing the
Fourier transform can be written as
\begin{eqnarray}
\frac{1}{\sqrt{N}}\sum_{j=0}^{N-1} \KET{j}\KET{f(j)}
&=&
\frac{1}{\sqrt{N}}\left\{
\sum_{j=0}^{M-1} \KET{j}\KET{f(j)}
+\sum_{j=M}^{2M-1} \KET{j}\KET{f(j)}
+\dots\right\}
\nonumber \\
&=&
\frac{1}{\sqrt{N}}
\sum_{j=0}^{M-1}
\left(\KET{j}+\KET{j+M}+\ldots\right) \KET{f(j)}
,
\end{eqnarray}
where, in the last step, we used the periodicity of $f(j)$.
Using the Fourier representation of $\KET{j}$ we obtain
\begin{eqnarray}
\frac{1}{\sqrt{N}}\sum_{j=0}^{N-1} \KET{j}\KET{f(j)}
&=&
\frac{1}{N}
\sum_{k=0}^{N-1}
\sum_{j=0}^{M-1}
e^{2\pi i k j/N}
\left(1+e^{2\pi i k M/N}+e^{4\pi i k M/N}+\ldots+e^{2\pi i k M(L-1)/N} \right)
\KET{k}\KET{f(j)}\nonumber\\
&+&
\frac{1}{N}
\sum_{k=0}^{N-1}
\sum_{j=0}^{L-1}
e^{2\pi i k j/N}e^{2\pi i k ML/N}
\KET{k}\KET{f(j)}
,
\end{eqnarray}
where $L=\lfloor N/M \rfloor$ denotes the largest integer $L$ such that $ML\le N$.
In simple terms, $L$ is the number of times the period $M$ fits into the
interval $[0,N-1]$.
The probability $p_q(M)$ to observe the quantum computer in the state $\KET{q}$ is given by
the expectation value of the (projection) operator $Q=\KET{q}\BRA{q}$.
With the restriction on $f(j)$ that $f(j)=f(j')$ implies $j=j'$,
we find
\begin{eqnarray}
\EXPECT{Q}=p_q(M)
&=&
\frac{M}{N^2}
\left(\frac{\sin(\pi qML/N)}{\sin(\pi qM/N)}\right)^2
+\frac{N-ML}{N^2}
\frac{\sin(\pi qM(2L+1)/N)}{\sin(\pi qM/N)}
.
\end{eqnarray}
The results for $p_q(M)$ in the case $N=8$ (three qubits) are given in Table~\ref{qfttab}.
From Table~\ref{qfttab} it follows directly that the expectation values
of the qubits are
($\EXPECT{Q_1^z}=\EXPECT{Q_2^z}=\EXPECT{Q_3^z}=0$) if the period $M=1$,
($\EXPECT{Q_1^z}=\EXPECT{Q_2^z}=0$, $\EXPECT{Q_3^z}=0.5$) if the period $M=2$,
($\EXPECT{Q_1^z}=0.5$, $\EXPECT{Q_2^z}=0.375$, $\EXPECT{Q_3^z}=0.34375$) if the period $M=3$,
and
($\EXPECT{Q_1^z}=0$, $\EXPECT{Q_2^z}=\EXPECT{Q_3^z}=0.5$) if the period $M=4$.

Thus, in this simple case of $N=15$, the periodicity of $f(j)$ can be unambiguously determined
from the expectation values of the individual qubits.
For $a=7$ we find ($\EXPECT{Q_1^z}=0$, $\EXPECT{Q_2^z}=0.5$, $\EXPECT{Q_3^z}=0.5$)
and hence the period $M=4$, yielding the correct factors g.c.d.$(7^2\pm1,15)=3,5$ of $N=15$.
Similarly, for $a=11$ we find ($\EXPECT{Q_1^z}=0$, $\EXPECT{Q_2^z}=0$, $\EXPECT{Q_3^z}=0.5$)
corresponding to the period $M=2$ and the factors g.c.d.$(11\pm1,15)=3,5$.

\begin{table}[t]
\caption{
Probability $p_q(M)$ to observe the state $\KET{q}$
after performing the quantum Fourier transform on the periodic function $f(j)=f(j+M)$
for $j=0,\dots,7$.}
\begin{center}
\begin{ruledtabular}
\begin{tabular}{ccccc}
$q$  & $p_q(M=1)$ & $p_q(M=2)$ & $p_q(M=3)$ & $p_q(M=4)$ \\
\hline
\noalign{\vskip 4pt}
 0  &  1  &  0.5  &  0.34375  & 0.25\\
 1  &  0  &  0.0  &  0.01451  & 0.00\\
 2  &  0  &  0.0  &  0.06250  & 0.25\\
 3  &  0  &  0.0  &  0.23549  & 0.00\\
 4  &  0  &  0.5  &  0.31250  & 0.25\\
 5  &  0  &  0.0  &  0.23549  & 0.00\\
 6  &  0  &  0.0  &  0.06250  & 0.25\\
 7  &  0  &  0.0  &  0.01451  & 0.00\\
\end{tabular}
\end{ruledtabular}
\label{qfttab}
\end{center}
\end{table}

\section{Event-based simulation of quantum phenomena}\label{sec3}

The conventional approach for simulating quantum systems
(and quantum computers in particular)
is to solve the time-dependent Schr\"odinger equation
of the corresponding system~\cite{QMvideo}.
In practice, this amounts to multiplying the state vector by a sequence
of (many) unitary matrices~\cite{RAED00,MICH03,RAED05}.
According to quantum theory, the squares of the amplitudes of the final state of the
quantum computer
yield the probabilities for observing the quantum computer in a particular basis state.
Once these probabilities are known, it is trivial to construct a random process that
generates events according to these probabilities.

The approach we propose in this paper is radically different.
We construct processes that generate events of which
the ratio of occurrence agrees with quantum theory.
Thus, adopting this method, we don't use concepts of quantum theory at all:
we don't use wave functions or the Schr\"odinger equation
and do not run into the fundamental measurement paradox~\cite{HOME97}.
In our approach, quantum mechanical behavior is the result
of a causal, deterministic (or stochastic), event-based process.

In quantum physics an event corresponds to the detection of a photon, electron, etc.
In our simulation approach an event is very much the same thing:
It is the arrival of a message at the input port of a processing unit.
In this paper we only consider networks of processing units in which
only one message is traveling through the network at any time.
Thus, the network receives an event at one of its inputs, processes
the event and delivers the processed message through one of its output channels.
After delivering this message the network can accept a new input event.

The key feature of our approach is a processing unit that
we call deterministic learning machine (DLM).
A \DLM\ is a machine with an internal state that
is updated according to a very simple, deterministic algorithm.
A \DLM\ responds to the input event by
choosing from all possible alternatives, the internal state
that minimizes the error between the input and the internal state itself.
This deterministic decision process determines
which type of event will be generated as output by the \DLM.
Furthermore, the same process generates different events in such a way that
the number of each type of event is proportional to the corresponding
probabilities of the quantum mechanical device that we want to simulate.
The message contains information about the decision the \DLM\ took
while updating its internal state and, depending on the application,
also contains other data that the \DLM\ can provide.
By updating its internal state, the \DLM\ ``learns" about the input events it receives
and by generating new events carrying messages, it tells
its environment about what it has learned.
This primitive learning capability is the essence of our approach.

\subsection{Deterministic learning machines}

An extensive discussion of the internal operation
and dynamic behavior of a \DLM\ can be found in Refs.~\cite{KOEN04,HANS05}.
Therefore, in this paper, we briefly recall the basic ideas.
As an example, we take a \DLM\ that we use to simulate
the Hadamard operation (see Section~\ref{sec2}).
We first consider a \DLM\ that accepts as input two different types
of events (0 or 1), each event carrying a message consisting of two real numbers.
The \DLM\ learns from the input event by updating its internal vector.
The internal state of the \DLM\ is represented by a unit vector of four real numbers
${\bf x}=(x_{0},x_{1},x_{2},x_{3})$.
The first (last) two elements of ${\bf x}$
are used to learn about the message carried by 0 (1) events.
As an input event is either of type 0 or 1,
the \DLM\ receives a message with two real numbers, not with four.
Therefore, the \DLM\ uses its internal
vector to supply the two missing real numbers.
Thus, for a 0 event carrying the message ${\bf y}_0=(y_{0},y_{1})$, the input vector
is ${\bf v}=(y_{0},y_{1},x_{2},x_{3})$ and
for a 1 event carrying the message ${\bf y}_1=(y_{2},y_{3})$, the input vector
is ${\bf v}=(x_{0},x_{1},y_{2},y_{3})$.
Consider now the second possibility in which the \DLM\
receives input in the form of four real numbers
${\bf v}=(v_{0},v_{1},v_{2},v_{3})$.
Then, it is clear that it can skip the step
of supplying the missing information and use the input ${\bf v}$ directly.

The \DLM\ updates its internal vector
by selecting from the eight candidate update rules $\{ j=0,1,2,3; s_j=\pm1\}$
\begin{eqnarray}
w_{i,j}=s_j\sqrt{1+\alpha^2(x_j^2-1)}\delta_{i,j}
+\alpha x_{i}(1-\delta_{i,j})
\label{HYP2}
\end{eqnarray}
the update rule that minimizes the cost
\begin{equation}
C = -{\bf w}_j^T{\bf v}^{\phantom{T}}.
\label{HYP1}
\end{equation}
Note that ${\bf x}^T{\bf x}^{\phantom{T}}=1$ implies
${\bf w}_j^T{\bf w}_j^{\phantom{T}}=1$ for each of the $8$ update rules.
The parameter $0<\alpha<1$ controls the learning process.
If $j^\prime$ and $s^\prime$ denote the values of $j$ and $s_j$
that minimizes the cost Eq.~\Eq{HYP1},
the final step of the \DLM\ algorithm is to set ${\bf x}={\bf w}_{j^\prime}$.

In general, the behavior of the \DLM\ defined by rules
Eqs.~\Eq{HYP2} and \Eq{HYP1} is difficult to analyze without the use of a computer.
However, for fixed input messages ${\bf y}_0=(y_{0},y_{1})$ and ${\bf y}_1=(y_{2},y_{3})$
for the 0 and 1 event respectively, it is clear what the \DLM\ will try to do:
It will minimize the cost Eq.~\Eq{HYP1} by rotating
its internal vector ${\bf x}$ to bring it as close as possible to
${\bf   y}=(y_{0},y_{1},y_{2},y_{3})$.
After a number of events (depending on the initial value of
${\bf x}$, the input ${\bf   y}$, and $\alpha$),
${\bf x}$ will be close to ${\bf   y}$.
However, the vector ${\bf x}$ does not converge to a limiting value
because the \DLM\ always changes its internal vector by a non-zero amount.
It is not difficult to see (and supported by simulations, results not shown)
that once ${\bf x}$ is close to ${\bf   y}$,
it will keep oscillating about ${\bf   y}=(  y_{0},  y_{1},  y_{2},  y_{3})$.

Let us denote by $n_0$ the number of times the
\DLM\ selects update rule $j=0$ (see Eq.\Eq{HYP2}).
Writing $w^2_{0,0}=(x_0+\delta)^2=1-\alpha^2+\alpha^2x^2_0$
and assuming that $0\ll\alpha<1$,
we find that the variable $x_0$ changes by an amount
$\delta\approx(1-\alpha^2)(1-x_0^2)/2x_0$
(neglecting terms of order $\delta^2$).
If $n$ is the total number of events then $n-n_0$ is the number of times
the \DLM\ selects update rules $j=1,2,3$.
For $j=1,2,3$ we have $w^2_{0,j}=(x_0+\delta^\prime)^2=\alpha^2x^2_0$
and hence $x_0$ changes by $\delta^\prime\approx-(1-\alpha^2)x_0/2$.
If ${\bf x}$ oscillates about ${\bf   y}$
then $x_0$ also oscillates about ${  y}_0$.
This implies that
the number of times $x_0$ increases times the increment must approximately be equal
to the number of times $x_0$ decreases times the decrement.
In other words, we must have $n_0\delta+(n-n_0)\delta^\prime\approx0$.
As $x_0\approx{  y}_0$ we conclude that $n_0/n\approx {  y}_0^2$.
Applying the same reasoning for the cases where
the \DLM\ selects update rule $j=1$ shows that
the number of times the \DLM\ will apply
update rules $j=0,1$ is proportional to ${  y}_0^2+{  y}_1^2$.
Therefore, the rate with which the \DLM\ selects update rules $j=0,1$
corresponds to the probability for observing a 0 event in the quantum mechanical system.
In other words, the \DLM\ generates 0 (1) events in a deterministic
manner, with a rate that is proportional to
the probability $p_0$ ($p_1=1-p_0$)
for observing a 0 (1) event in the corresponding quantum mechanical system.
Thus, a \DLM\ is a simple ``classical'' dynamical system that exhibits
behavior that is usually attributed to quantum systems.

\setlength{\unitlength}{1cm}
\begin{figure*}[t]
\begin{center}
\setlength{\unitlength}{1cm}
\begin{picture}(14,6.5)
\put(-1.75,0){\includegraphics[width=9cm]{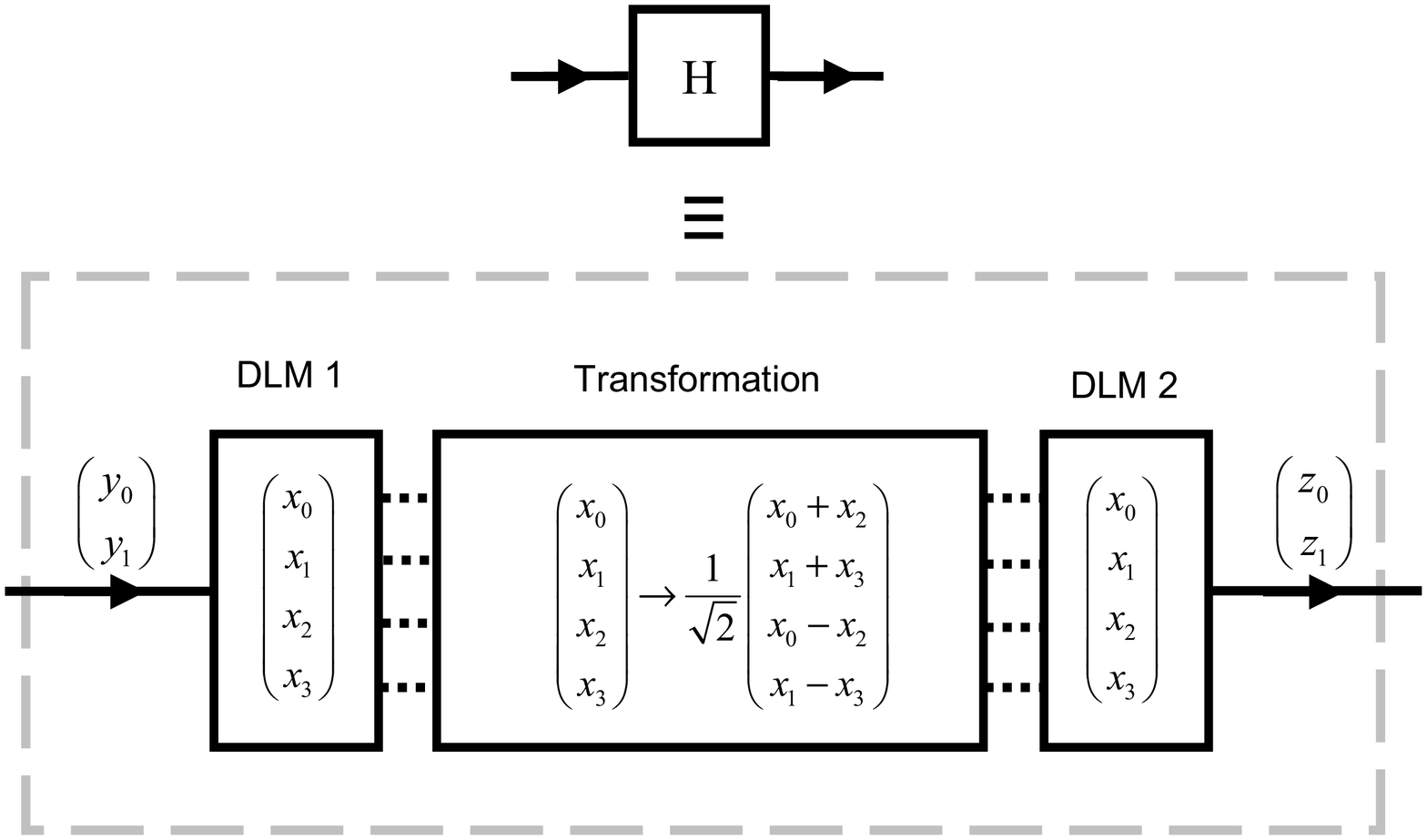}}
\put(7.3,0){\includegraphics[width=9cm]{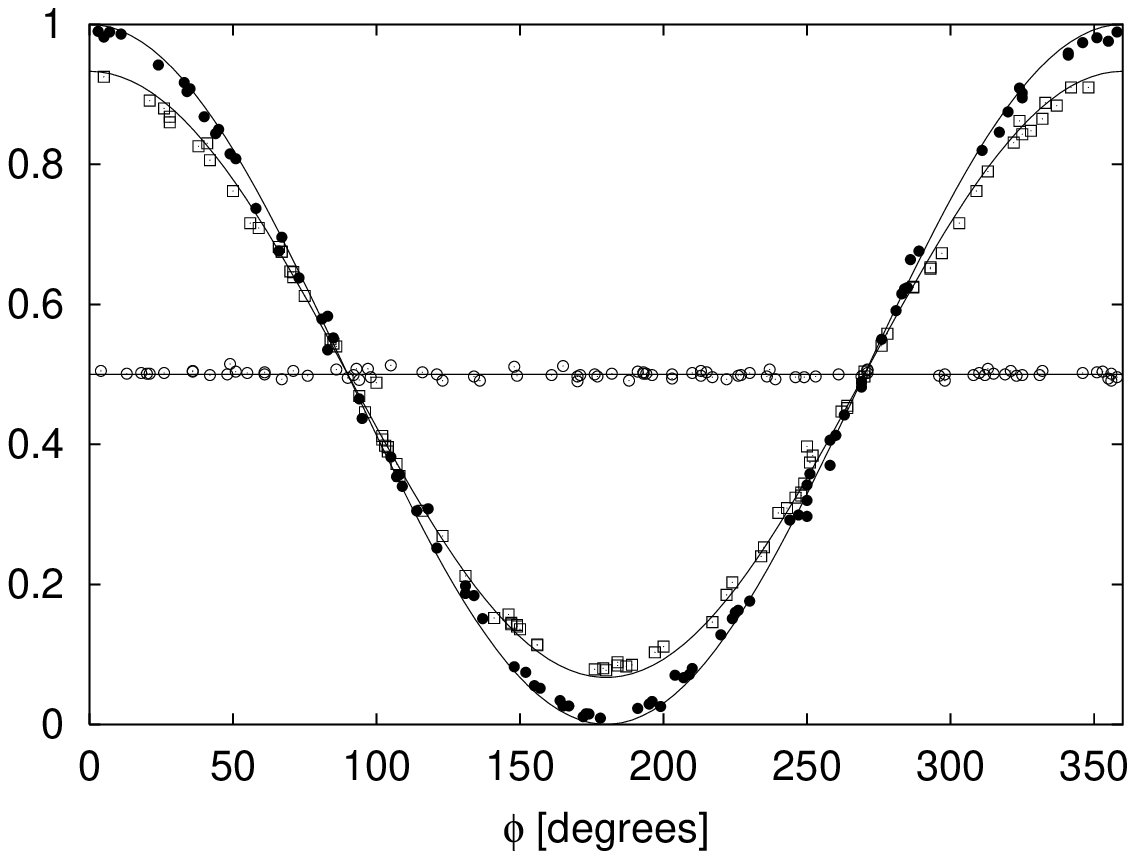}}
\end{picture}
\caption{
Left:
Diagram of the network of two \DLMS\ that performs a deterministic
simulation of a Hadamard gate on an event-by-event basis.
The arrows on the solid lines represent the input and output events.
Dashed lines indicate the flow of data within the \DLM-based processor.
Right:
Simulation results for the Hadamard gate shown on the left.
The input events are either of type
0 with message $(\cos\psi_0,\sin\psi_0)$
or of type 1 with message $(\cos\psi_1,\sin\psi_1)$.
A uniform random number is used to generate the type of input events.
The probability for a 0 (1) event is $p_0$ ($p_1=1-p_0$).
Each data point represents a simulation of 10000 events.
After each set of 10000 events, a uniform
random number in the range $[0,360]$ is used to
choose the angles $\psi_0$ and $\psi_1$.
Markers give the simulation results for the
normalized intensity in output channel 0 as a function of
$\phi=\psi_0-\psi_1$.
Open circles: $p_0=1$;
Bullets: $p_0=0.5$;
Open squares: $p_0=0.25$.
Lines represent the results of quantum theory (see Eq.\Eq{BS4}).
}
\label{figbs}
\label{one-bs}
\label{hadamard}
\end{center}
\end{figure*}

\subsection{Stochastic learning machines}\label{SLM}

The sequence of events that is generated by a \DLM\ (network) is strictly deterministic.
We now describe a simple modification that
turns a \DLM\ into a stochastic learning machine (SLM).
The term {\sl stochastic} does not refer to the learning process but to
the method that is used to select the output channel that carries the outgoing
message.

In the stationary regime, the components of the internal vector represent
the probability amplitudes.
Comparing the (sums of) squares of these amplitudes with a uniform
random number $0<r<1$ gives the probability for sending the message
over the corresponding output channel.
For instance, in the case of the Hadamard gate (see Fig.~\ref{hadamard})
we replace the \DLM\ 2 by a SLM.
This SLM generates a 0 event if $x^2_0+x^2_1\le r$
and a 1 event otherwise.
Although the learning process of this processor is still
deterministic, in the stationary regime the output events are
randomly distributed over the two possibilities.
Of course, the rate at which different output events are generated
is the same as that of the original \DLM-network.
Replacing \DLMS\ by SLMs in a \DLM-network changes the order in which
messages are being processed by the network but leaves the content of the
messages intact.
Therefore, in the stationary regime, the distribution of messages
over the outputs of the SLM-network is essentially the same as that
of the original DLM network.

\section{Event-based simulation of quantum computers}\label{sec4}

\subsection{Hadamard operation~\cite{MZIdemo}}\label{HADA}

As an example of a \DLM-based processor that performs single-qubit operations
we consider the diagram shown in Fig.~\ref{one-bs} (left).
The presence of a message is indicated by an arrow on the corresponding line.
The first component, called front-end,
consists of a \DLM\ that ``learns'' about the occurrence of 0 and 1 events,
meaning that the corresponding qubit is 0 or 1, respectively.
The second component transforms the data stored in the front-end
and feeds this data into a second \DLM\, called back-end.
The back-end ``learns'' this data.
The learning process itself is used to determine whether
the back-end responds to the input event by sending out either a 0 or a 1 event.
None of these components makes use of random numbers.

An event corresponds to the arrival of a particle in either the 0 or 1 state.
The message is a unit vector ${\bf y}=(y_{0},y_{1})$ of two real numbers.
We denote the number of 0 (1) events by $N_0$ ($N_1$) and
the total number of events by $N=N_0+N_1$.
The correspondence with the quantum system is rather obvious:
the probability for a 0 event is given by
$|a_0|^2\approx N_0/N$ and $y_{0}=\hbox{Re } a_0/|a_0|$ and $y_{1}=\hbox{Im } a_0/|a_0|$.
The probability for a 1 event is
$N_1/N\approx|a_1|^2$ and $y_{2}=\hbox{Re } a_1/|a_1|$ and $y_{3}=\hbox{Im } a_1/|a_1|$.

From Fig.~\ref{one-bs} (left) it is clear that
the transformation is just the real-valued version
of the complex-valued matrix-vector operation
that corresponds to the Hadamard gate (see Eq.\Eq{SPIN10}).
A processor that performs
the general single-qubit operation Eq.\Eq{SPIN11}
is identical to the one shown in Fig.~\ref{one-bs} (left)
except for the transformation stage.
For instance, to implement the $X$ operation
(see Eq.\Eq{SPIN10}) we only have to replace the
transformation matrix of the Hadamard operation
\begin{equation}
\frac{1}{\sqrt{2}}
\left(
\begin{array}{cccc}
\phantom{-}1&\phantom{-}0 &\phantom{-}1 &\phantom{-}0 \\
\phantom{-}0&\phantom{-}1 &\phantom{-}0 &\phantom{-}1 \\
\phantom{-}1&\phantom{-}0 &-1 &\phantom{-}0 \\
\phantom{-}0&\phantom{-}1 &\phantom{-}0 &-1 \\
\end{array}
\right)
\Longleftrightarrow
H
,
\label{HADA1}
\end{equation}
by
\begin{equation}
\frac{1}{\sqrt{2}}
\left(
\begin{array}{cccc}
\phantom{-}1&\phantom{-}0 &\phantom{-}0 &          -1 \\
\phantom{-}0&\phantom{-}1 &\phantom{-}1 &\phantom{-}0 \\
\phantom{-}0&          -1 &\phantom{-}1 &\phantom{-}0 \\
\phantom{-}1&\phantom{-}0 &\phantom{-}0 &\phantom{-}1 \\
\end{array}
\right)
\Longleftrightarrow
X
.
\label{HADA2}
\end{equation}

In Fig.~\ref{one-bs} (right) we present results of simulations using
the processor depicted in Fig.~\ref{figbs} (left).
Before the first simulation starts we
use uniform random numbers to initialize the
two four-dimensional internal vectors of DLM 1 and DLM 2.
Each data point in Fig.~\ref{one-bs} (right) represents a simulation of 10000 events.
All these simulations were carried out with $\alpha=0.99$.
For each set of 10000 events, two uniform
random numbers in the range $[0,360]$ determine two angles $\psi_0$ and $\psi_1$
that are used as the message ${\bf y}_0=(\cos\psi_0,\sin\psi_0)$
(${\bf y}_1=(\cos\psi_1,\sin\psi_1)$)
for the input event of type 0 (1).
Uniform random numbers are used to generate 0 (1) input events
with probability $p_0$ ($p_1=1-p_0$).
This corresponds to the input amplitudes
$a_0=\sqrt{p_0}e^{i\psi_0}$ and $a_1=p_1 e^{i\psi_1}$
in the quantum mechanical system.

According to quantum theory, the probability amplitude $b_0$ ($b_1$)
for the 0 (1) output event is given by
\begin{eqnarray}
\left(
\begin{array}{c}
b_0\\
b_1
\end{array}
\right)
=
\frac{1}{\sqrt{2}}
\left(
\begin{array}{c}
a_0+a_1\\
a_1-a_0
\end{array}
\right)
=
\frac{1}{\sqrt{2}}
\left(
\begin{array}{cc}
\phantom{-}1&\phantom{-}1\\
\phantom{-}1&-1
\end{array}
\right)
\left(
\begin{array}{c}
a_0\\
a_1
\end{array}
\right).
\label{BS3}
\end{eqnarray}
As the \DLM-based Hadamard gate generates
$N_0/N$ events of type 0 and
$N_1/N$ events of type 1,
it is obvious from Fig.~\ref{one-bs} (right) that these ratios
are in excellent agreement with the probabilities
\begin{eqnarray}
|b_0|^2&=&\frac{1+2\sqrt{p_0(1-p_0)}\cos(\psi_0-\psi_1)}{2},\nonumber \\
|b_1|^2&=&\frac{1-2\sqrt{p_0(1-p_0)}\cos(\psi_0-\psi_1)}{2},
\label{BS4}
\end{eqnarray}
as obtained from Eq.\Eq{BS3}.

\setlength{\unitlength}{1cm}
\begin{figure*}[t]
\begin{center}
\setlength{\unitlength}{1cm}
\includegraphics[width=12cm]{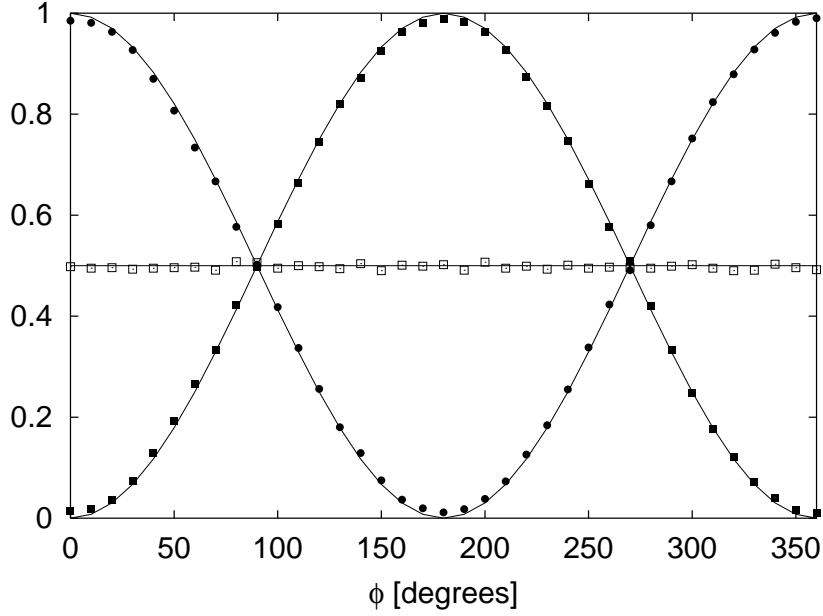}
\caption{
Simulation results for the \DLM-implementation of the network shown in Fig.~\ref{one-mzi}.
The input are 0 events with message $(\cos\psi_0,\sin\psi_0)$.
A uniform random number in the range $[0,360]$ is used to choose the angle $\psi_0$.
Each data point represents 10000 events ($N_0+N_1=N_2+N_3=10000$).
Initially the rotation angle $\phi=0$ and after each set of 10000 events, $\phi$
is increased by $10^\circ$.
Markers give the simulation results for the normalized intensities
as a function of $\phi$.
Open squares: $N_0/(N_0+N_1)$ (0 events);
Solid squares: $N_2/(N_2+N_3)$ (0 events);
Bullets: $N_3/(N_2+N_3)$ (1 events).
Lines represent the results of quantum theory.
}
\label{twox}
\end{center}
\end{figure*}

\subsection{Mach-Zehnder interferometer~\cite{MZIdemo}}\label{MZI}

As a second example of event-based simulation of single-qubit operations
we consider the Mach-Zehnder interferometer
network shown in Fig.~\ref{one-mzi}.
We use the equivalent \DLM-based processor for the $X$ operation.
The phase-shift operation $R(\phi)$ is carried out
by a passive device (that is, a device without \DLMS) that
simply passes messages of 0 events and transforms
messages of 1 events by performing a plane
rotation about $\phi$ of the two-dimensional vector representing the message.
Thus, $R(\phi)$ transforms
the message ${\bf y}=(y_{0},y_{1})$ carried by a 1 event according to
\begin{eqnarray}
\left(
\begin{array}{c}
y_0\\
y_1
\end{array}
\right)
\longleftarrow
\frac{1}{\sqrt{2}}
\left(
\begin{array}{cc}
\phantom{-}\cos\phi&-\sin\phi\\
\phantom{-}\sin\phi&\phantom{-}\cos\phi
\end{array}
\right)
\left(
\begin{array}{c}
y_0\\
y_1
\end{array}
\right).
\label{MZI1}
\end{eqnarray}

In Fig.~\ref{twox} we present a few typical simulation results
for the Mach-Zehnder interferometer built from \DLMS.
We assume that the input receives 0 events only,
each event carrying the message $(\cos\psi_0,\sin\psi_0)$.
This corresponds to $(a_0,a_1)=(\cos\psi_0+i\sin\psi_0,0)$
in the quantum system.
We use a uniform random number to determine $\psi_0$.
In all these simulations $\alpha=0.99$.
The number of 0 (1) events after the
first $X$ operation is denoted by $N_0$ ($N_1$).
The number of 0 (1) events after the
last $X$ operation is denoted by $N_2$ ($N_3$).
The data points in Fig.~\ref{twox} are the simulation results for the
normalized intensity $N_i/(N_0+N_1)$ for $i=0,2,3$ as a function of $\phi$.
Lines represent the corresponding results of quantum theory (see Eq.\Eq{MZ3}).
From Fig.~\ref{twox} it is clear
that the event-based processing by the \DLM\ network reproduces
the probability distribution as obtained from Eq.\Eq{MZ3}.
Messages generated by \DLMS\ preserve the phase information
that is essential for the system to exhibit quantum interference effects.

Summarizing: We have shown that a \DLM-network can simulate
single-photon quantum interference particle-by-particle
without using quantum theory.
In practicular, the previous example demonstrates
that locally-connected networks of processing units
with a primitive learning capability are sufficient
to simulate, event-by-event, the single-photon beam splitter and Mach-Zehnder interferometer
experiments of Grangier et al.~\cite{GRAN86}.
The parts of the processing units and network map one-to-one
on the physical parts of the experimental setup
and only simple geometry is used to construct the simulation algorithm.
In this sense, the simulation approach we propose
satisfies Einstein's criteria of realism and causality~\cite{HOME97}.

\subsection{CNOT operation}\label{CNOTsec}

\begin{figure*}[t]
\begin{center}
\setlength{\unitlength}{1cm}
\includegraphics[width=14cm]{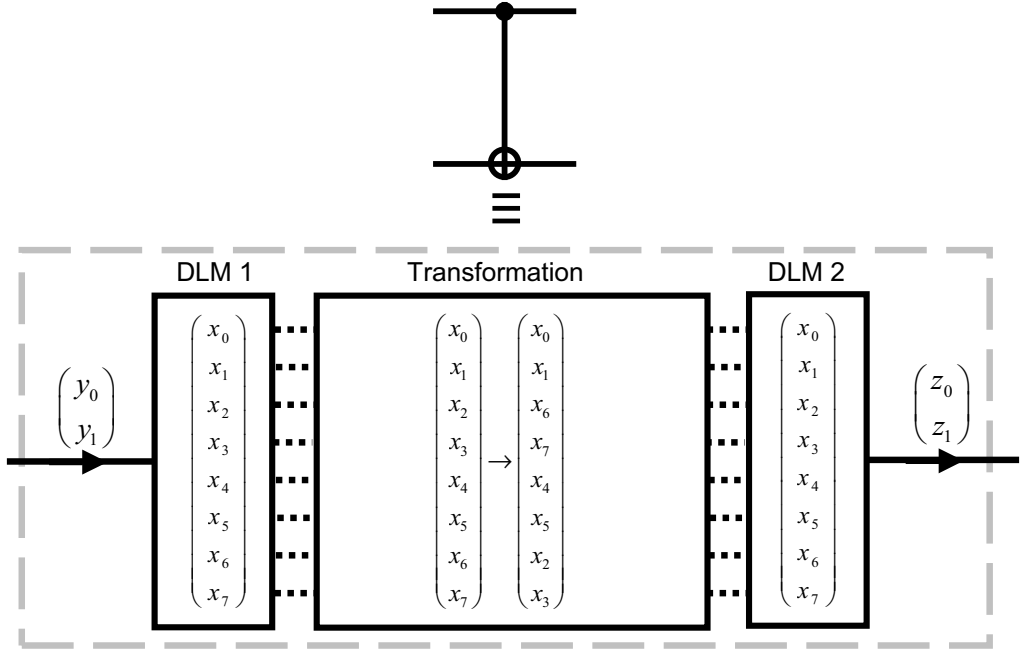}
\caption{
Diagram of a \DLM-based processor that simulates a CNOT gate
on an event-by-event basis.
}
\label{figcnot}
\end{center}
\end{figure*}

The schematic diagram
of the \DLM-network that performs the CNOT operation on an
event-by-event (particle-by-particle) basis is shown in Fig.~\ref{figcnot}.
Conceptually the structure of this network
is the same as in the case of a system of a single qubit.
As input to the \DLM-network we now have four (0,1,2 or 3)
instead of two different types of events,
corresponding to the quantum states
$\KET{00}$,
$\KET{01}$,
$\KET{10}$,
$\KET{11}$.
Each event carries a message consisting of two real numbers
${\bf y}=(\cos\phi_i,\sin\phi_i)$ for
$i=0,\ldots,3$, corresponding to the phase of the quantum
mechanical probability amplitudes, that is  $a_0/|a_0|,\ldots, a_3/|a_3|$.
The internal state of each \DLM\ is represented by a unit vector of eight real numbers
${\bf x}=(x_{0},\dots,x_{7})$ and there are
sixteen candidate update rules ($\{ j=0,\ldots7; s_j=\pm1\}$, see Eq.\Eq{HYP2})
to choose from.
The rule that is actually used is determined by minimizing the
cost function $C = -s_j{\bf w}_j^T{\bf v}^{\phantom{T}}$.
The transformation stage is extremely simple:
According to Eq.\Eq{CNOT2}, all it has to do is swap the two pairs
of elements ($x_2$,$x_3$) and ($x_6$,$x_7$).

\setlength{\unitlength}{1cm}

Instead of presenting results that show that the \DLM-processor
of Fig.~\ref{figcnot} correctly performs the CNOT operation
on an event-by-event basis,
we consider the more complicated
network of four Hadamard gates and one CNOT gate shown in Fig.~\ref{figcnot2}.
Quantum mechanically, this network acts as a CNOT gate
in which the role of control- and target qubit have been interchanged~\cite{NIEL00}.
For this \DLM-network to perform correctly it is essential that
the event-based simulation mimics the quantum interference (generated by the Hadamard
gates) correctly.
In Table~\ref{CNOTdata} we present simulation results for the \DLM-network shown in
Fig.~\ref{figcnot2}.
Before the simulation starts, we
use uniform random numbers to initialize the
internal vectors of the \DLMS\ (ten vectors in total).
For simplicity, we take as input messages $\phi_0=\phi_1=\phi_2=\phi_3=0$.
All these simulations were carried out with $\alpha=0.99$.
From Table~\ref{CNOTdata} it is clear that, also for a modest number
of input events, the network reproduces
the results of the corresponding quantum circuit, i.e. a CNOT operation
in which qubit 2 is the control qubit and qubit 1 is the target qubit~\cite{NIEL00}.

\begin{table}[t]
\caption{Simulation results for the \DLM-network shown in
Fig.~\ref{figcnot2}, demonstrating that the network reproduces
the results of the corresponding quantum circuit, i.e. a CNOT operation
in which qubit 2 is the control qubit and qubit 1 is the target qubit~\cite{NIEL00}.
The first half of the events are discarded in the calculation
of the frequency $f_i$ for observing an output event
of type $i=0,1,2,3$, corresponding to the probability to
observe the quantum system in the state $\KET{00}$,
$\KET{01}$, $\KET{10}$, or $\KET{11}$, respectively.
For 200 events or more, the difference between the event-based simulation results
and the corresponding quantum mechanical probabilities is less than 1\%.
}
  \begin{center}
    \begin{ruledtabular}
    \begin{tabular}{ccccccc}
Number of events&Qubit 1 &Qubit 2&$f_0$&$f_1$&$f_2$&$f_3$\\
\hline
\noalign{\vskip 4pt}
100&0&0&$0.98$&$0.00$&$0.00$&$0.02$\\ 
100&1&0&$0.20$&$0.74$&$0.01$&$0.04$\\ 
100&0&1&$0.16$&$0.04$&$0.00$&$0.80$\\ 
100&1&1&$0.16$&$0.04$&$0.72$&$0.08$\\ 
\noalign{\vskip 2pt}
\hline
\noalign{\vskip 2pt}
200&0&0&$1.00$&$0.00$&$0.00$&$0.00$\\
200&1&0&$0.00$&$1.00$&$0.00$&$0.00$\\
200&0&1&$0.00$&$0.00$&$0.00$&$1.00$\\
200&1&1&$0.01$&$0.00$&$0.99$&$0.00$\\
    \end{tabular}
    \end{ruledtabular}
    \label{CNOTdata}
  \end{center}
\end{table}

As an illustration of the use of SLMs, we replace all the back-end \DLMS\
in the CNOT circuit shown in Fig.~\ref{figcnot} by SLMs.
and repeat the simulations that yield the data in Table~\ref{CNOTdata}.
From Table~\ref{CNOTdata2} we conclude that the randomized
version generates the correct results but significantly more events
are needed to achieve similar accuracy as in the fully deterministic
simulation.

\begin{table}[t]
\caption{Simulation results for the \DLM-network shown in
Fig.~\ref{figcnot2} in which each back-end \DLM\ of the individual
gates has been replaced by a SLM.
The latter uses random numbers to randomize the order in
which different output events are generated but does not change
the frequencies of the events.
The first half of the input events are discarded for the calculation
of the frequencies $f_i$ for observing an output event
of type $i=0,1,2,3$, corresponding to the probability to
observe the quantum system in the state $\KET{00}$,
$\KET{01}$, $\KET{10}$, or $\KET{11}$, respectively.}
  \begin{center}
    \begin{ruledtabular}
    \begin{tabular}{ccccccc}
Number of events&Qubit 1 &Qubit 2&$f_0$&$f_1$&$f_2$&$f_3$\\
\hline
\noalign{\vskip 4pt}
2000&0&0&$0.965$&$0.015$&$0.010$&$0.010$\\
2000&1&0&$0.007$&$0.970$&$0.012$&$0.011$\\
2000&0&1&$0.010$&$0.008$&$0.016$&$0.966$\\
2000&1&1&$0.005$&$0.016$&$0.963$&$0.016$\\
    \end{tabular}
    \end{ruledtabular}
    \label{CNOTdata2}
  \end{center}
\end{table}

\subsection{Number factoring}

Finally, we discuss the results obtained by a \DLM-based simulation
of the number factoring circuits depicted in Fig.~\ref{shor-circuit}.
\DLM\ networks (not shown) that perform the Toffoli gate operation and the
Fourier transform, which involves several Hadamard
operations and controlled phase shifts, are readily
constructed by mimicking the procedure for the
construction of the \DLM\ network of the CNOT gate.
Having done this, building the circuit in Fig.~\ref{shor-circuit}
entails nothing than connecting the \DLM\ networks that
simulate the various quantum gates.
As this circuit involves seven qubits,
the internal vectors of the \DLMS\ have 256 elements.

Section~\ref{SHOR} shows that
if $a=7$, we expect to find
$Q_1=\EXPECT{Q_1^z}=0$, $Q_2=\EXPECT{Q_2^z}=0.5$, and $Q_3=\EXPECT{Q_3^z}=0.5$.
Similarly, for $a=11$ we expect to find
$Q_1=\EXPECT{Q_1^z}=0$, $Q_2=\EXPECT{Q_2^z}=0$, and $Q_3=\EXPECT{Q_3^z}=0.5$.
In the \DLM\ approach, simply counting the number of 1 events in the three output channels of
the Fourier transform (inside the dashed box in Fig.~\ref{shor-circuit})
and dividing these numbers by the total number of events analyzed,
we obtain numerical estimates for the qubits
$Q_1=\EXPECT{Q_1^z}$, $Q_2=\EXPECT{Q_2^z}$, and $Q_3=\EXPECT{Q_3^z}$.
In Figs.~\ref{figa7-a}-\ref{figa11} we present simulation results for the
\DLM\ (left panel) and SLM (right panel)
implementation of the circuit depicted in Fig.~\ref{shor-circuit},
for the two cases $a=7,11$ and for two choices of the control
parameter $\alpha=0.99,0.999$.
After processing a few events, (less than 200 if $\alpha=0.99$
less than 2000 if $\alpha=0.999$) all the \DLM\ networks
reproduce the results of quantum theory with high accuracy.
Replacing \DLMS\ by their stochastic equivalents (SLMs), we
see that the fluctuations are larger than if we use \DLMS\ and
that the system as a whole ``learns'' much slower.
This is to be expected: the probabilistic mechanism
to distribute events over the output channels of the SLMs
makes much more ``mistakes'' than the deterministic process
used by the \DLMS.
\DLMS\ encode the information about the probability and phase
in a much more effective, compact manner than SLMs.
In the case of the latter, the correct probability distribution
is encoded in a statistical manner
and can only be recovered by analyzing a lot of events.

From the description of the learning process, it is
clear that $\alpha$ controls the rate of learning or, equivalently,
the rate at which learned information can be forgotten.
Furthermore it is evident that the difference between
a constant input to a \DLM\ and the learned value of its
internal variable cannot be smaller than $1-\alpha$.
In other words, $\alpha$ also limits the precision with
which the internal variable can represent a sequence
of constant input values.
On the other hand, the number of events has to balance
the rate at which the \DLM\ can forget a learned input value.
The smaller $1-\alpha$ is, the larger the number of events
has to be for the \DLM\ to adapt to changes in the input data.
The results depicted in Figs.~\ref{figa7-a}-\ref{figa11-a}
and
Figs.~\ref{figa7}-\ref{figa11}
confirm this behavior.

\setlength{\unitlength}{1cm}
\begin{figure*}[t]
\begin{center}
\setlength{\unitlength}{1cm}
\begin{picture}(14,6.5)
\put(-2.,0){\includegraphics[width=9cm]{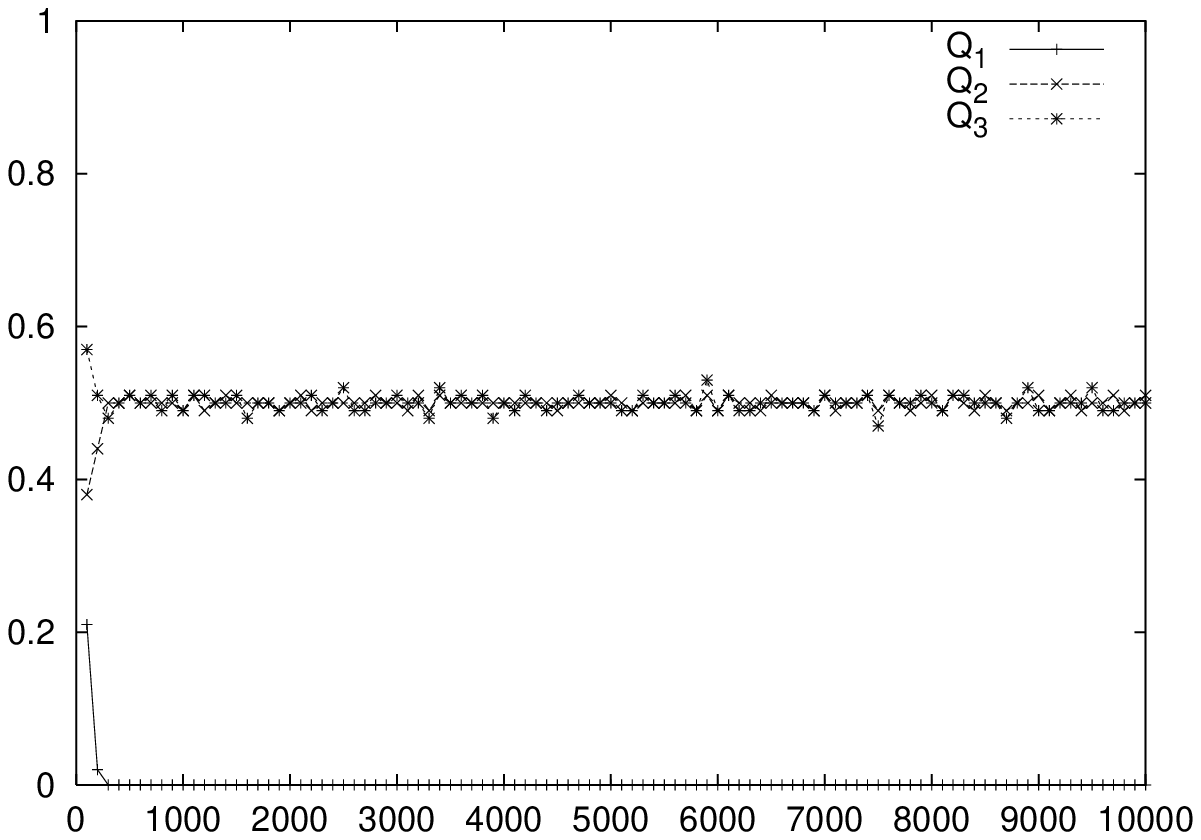}}
\put(7.,0){\includegraphics[width=9cm]{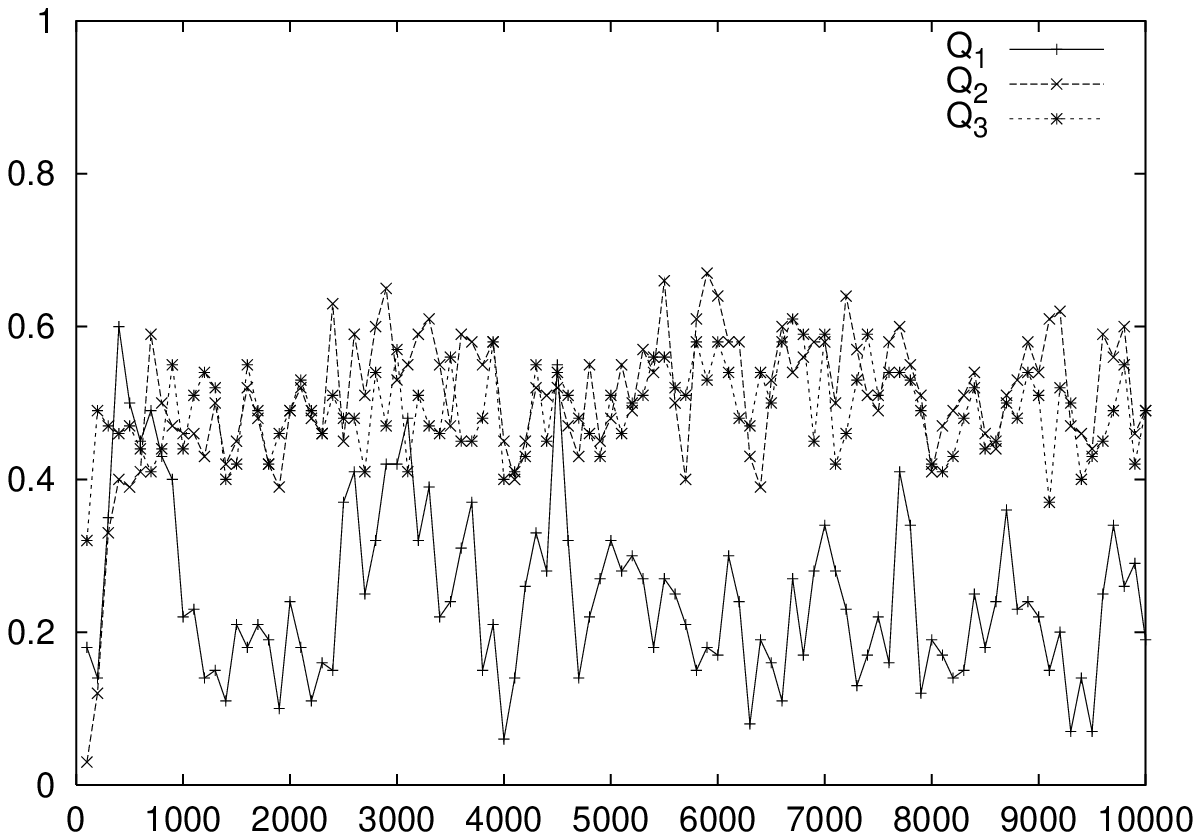}}
\end{picture}
\caption{
Event-by-event simulation of Shor's quantum algorithm for factoring
the integer $N=15$, using the value $a=7$ (see Section~\ref{SHOR}).
Each data point represents the average of 100 output events.
The parameter that controls the learning process
of the \DLMS\ is $\alpha=0.99$.
Left: Deterministic simulation employing \DLMS.
Right: Stochastic simulation employing SLMs.
}
\label{figa7-a}
\end{center}
\end{figure*}
\setlength{\unitlength}{1cm}
\begin{figure*}[t]
\begin{center}
\setlength{\unitlength}{1cm}
\begin{picture}(14,6.5)
\put(-2.,0){\includegraphics[width=9cm]{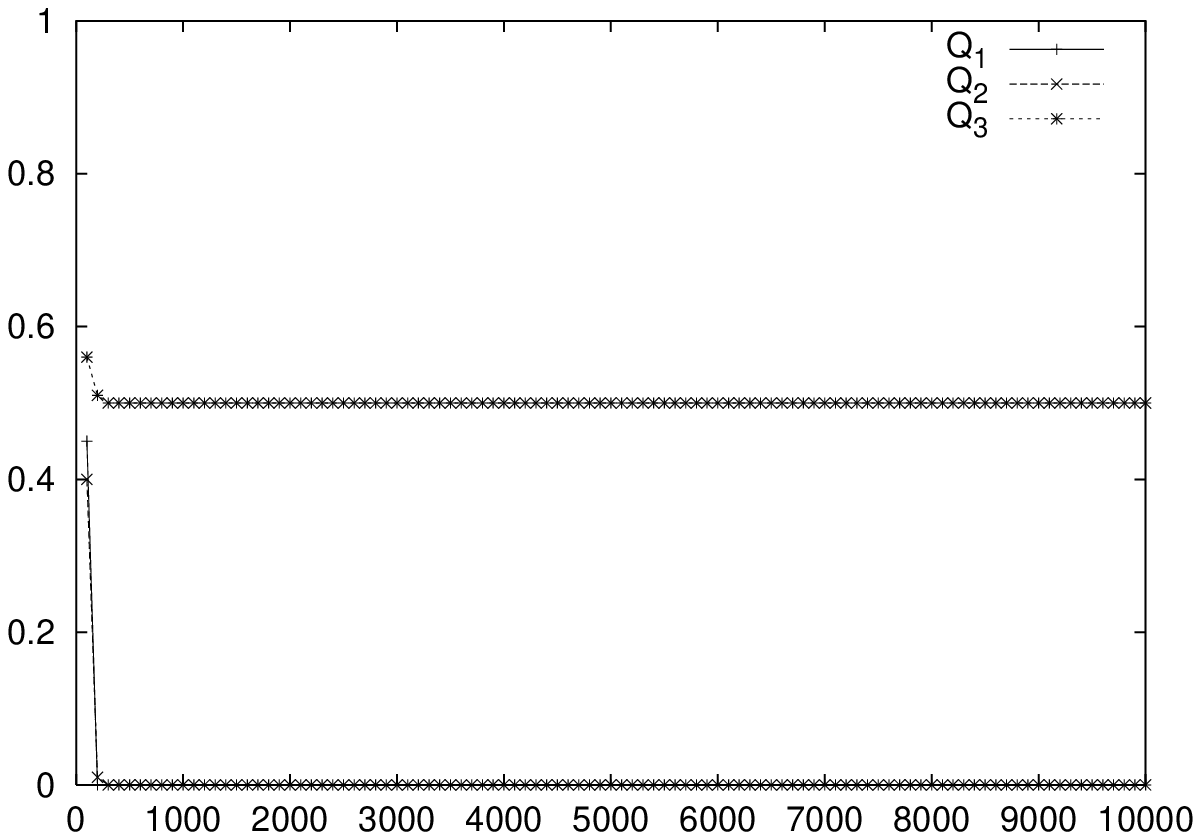}}
\put(7.,0){\includegraphics[width=9cm]{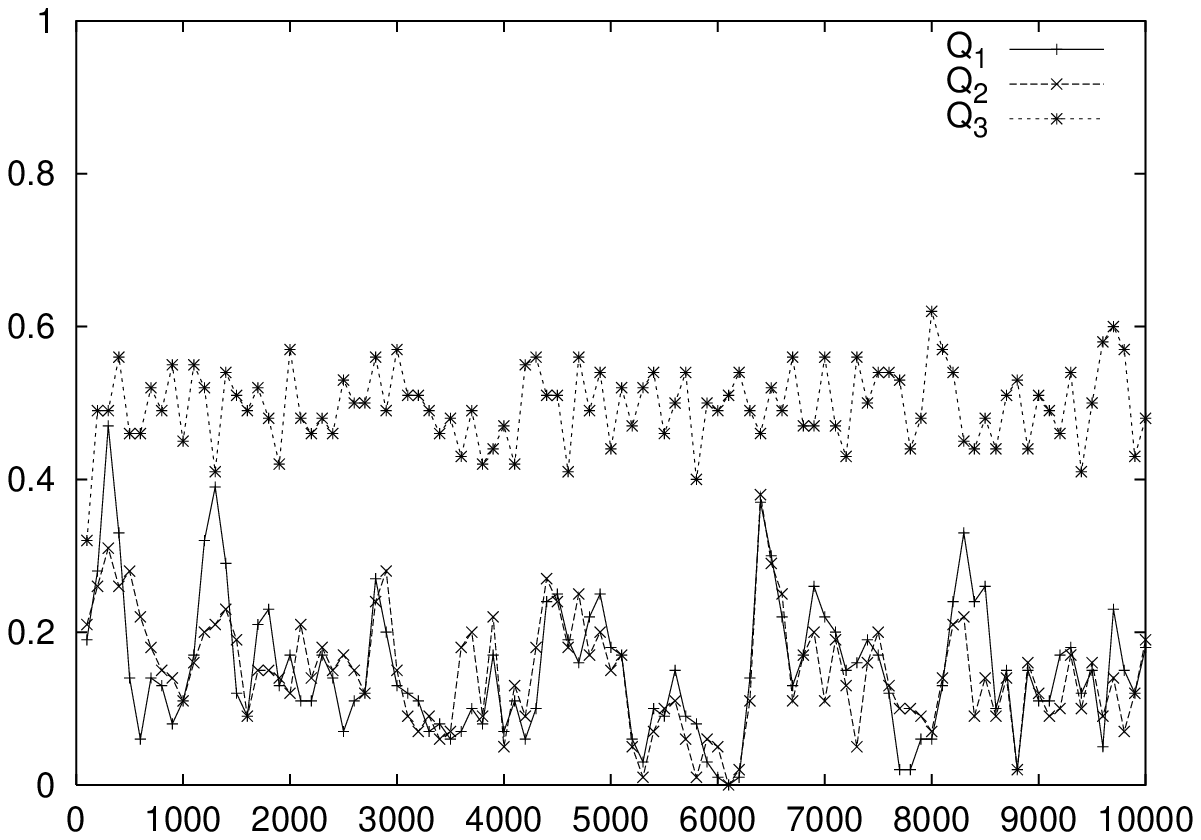}}
\end{picture}
\caption{
Same as Fig.~\ref{figa7-a}, except that $a=11$ (see Section~\ref{SHOR}).
}
\label{figa11-a}
\end{center}
\end{figure*}
\setlength{\unitlength}{1cm}
\begin{figure*}[t]
\begin{center}
\setlength{\unitlength}{1cm}
\begin{picture}(14,6.5)
\put(-2.,0){\includegraphics[width=9cm]{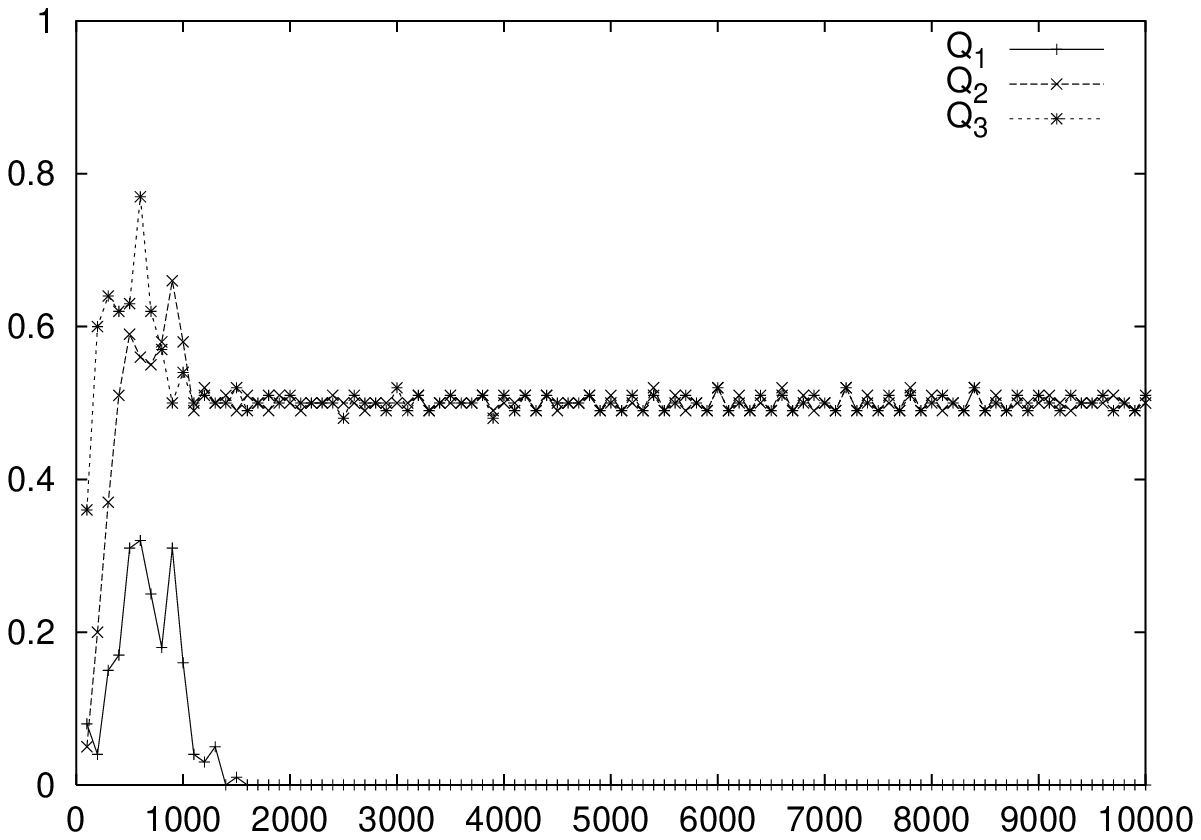}}
\put(7.,0){\includegraphics[width=9cm]{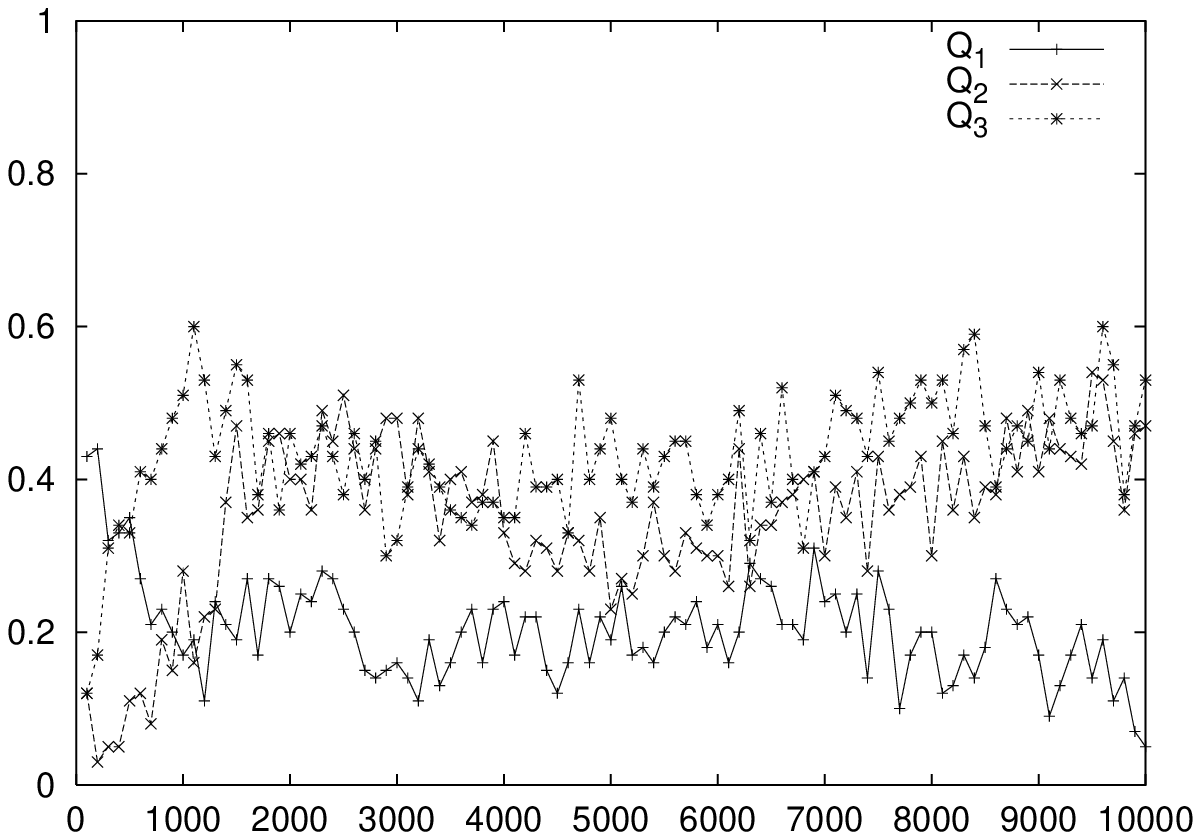}}
\end{picture}
\caption{
Event-by-event simulation of Shor's quantum algorithm for factoring
the integer $N=15$, using the value $a=7$ (see Section~\ref{SHOR}).
Each data point represents the average of 100 output events.
The parameter that controls the learning process
of the \DLMS\ is $\alpha=0.999$.
Left: Deterministic simulation employing \DLMS.
Right: Stochastic simulation employing SLMs.
}
\label{figa7}
\end{center}
\end{figure*}
\setlength{\unitlength}{1cm}
\begin{figure*}[t]
\begin{center}
\setlength{\unitlength}{1cm}
\begin{picture}(14,6.5)
\put(-2.,0){\includegraphics[width=9cm]{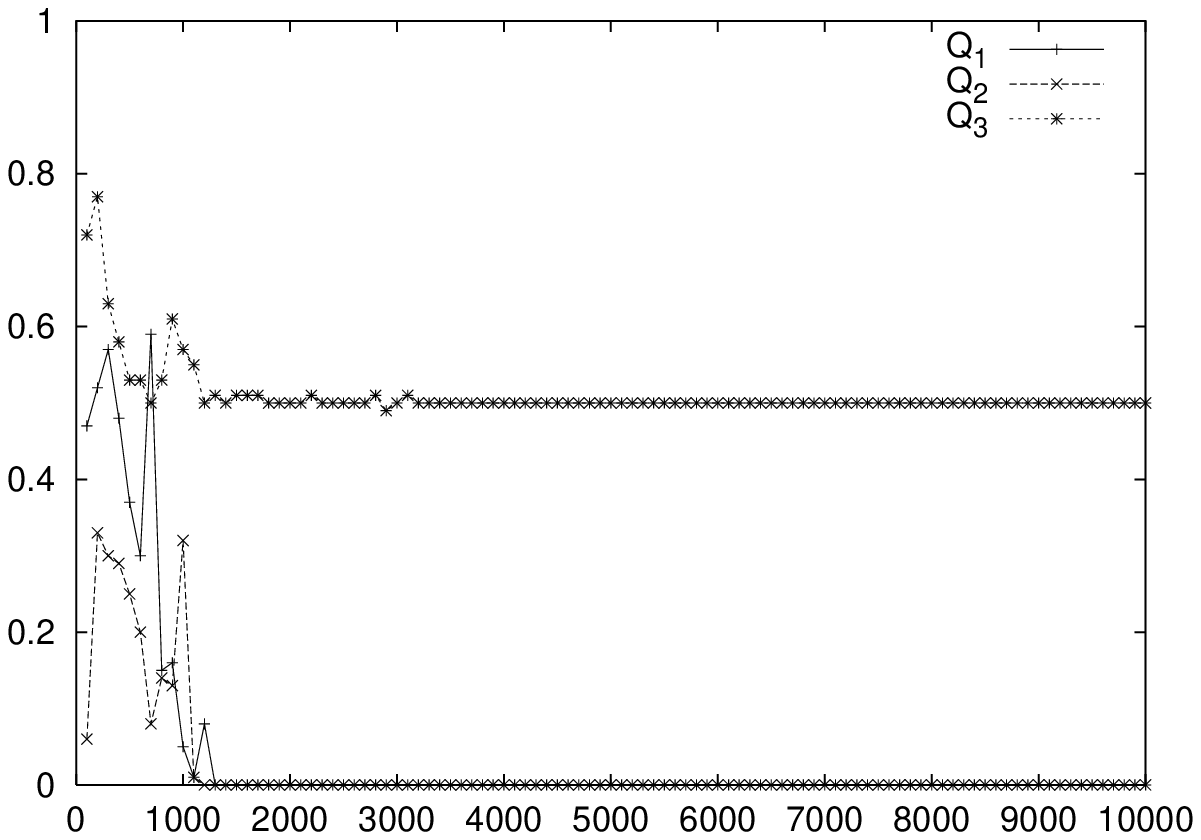}}
\put(7.,0){\includegraphics[width=9cm]{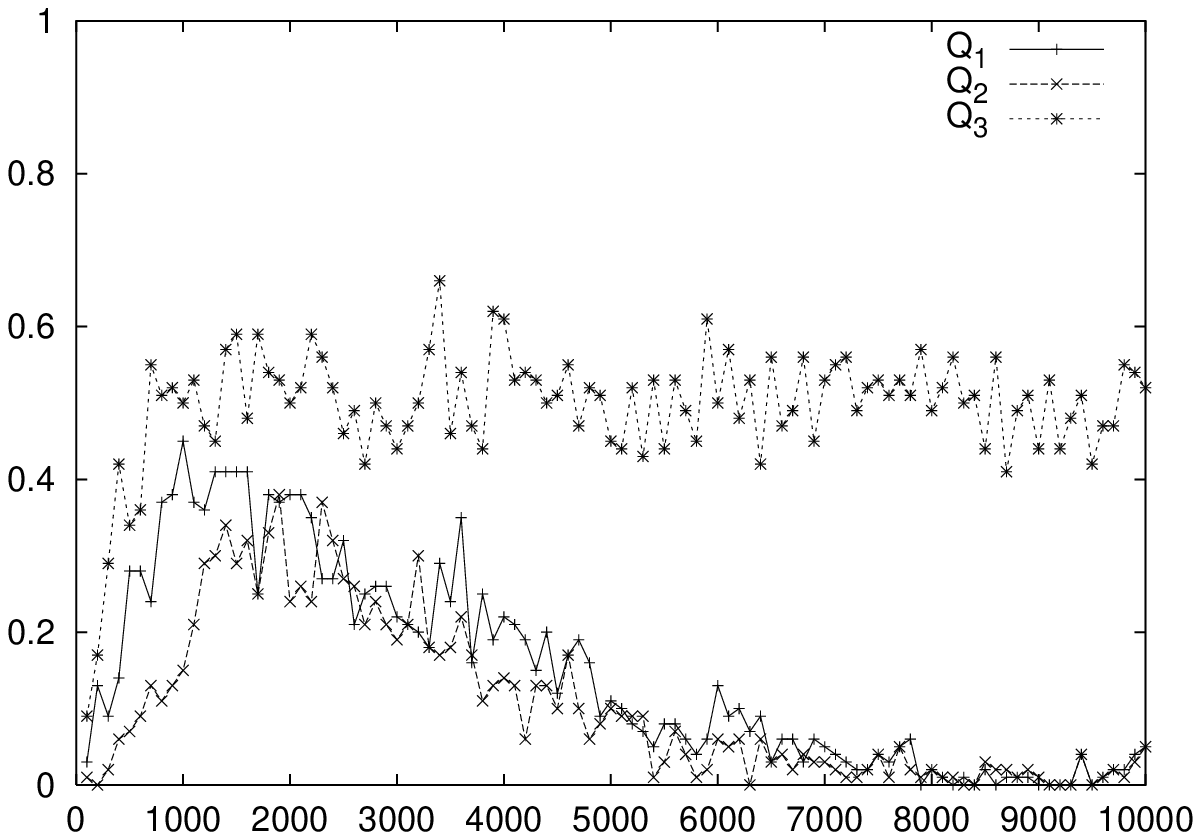}}
\end{picture}
\caption{
Same as Fig.~\ref{figa7}, except that $a=11$ (see Section~\ref{SHOR}).
}
\label{figa11}
\end{center}
\end{figure*}

\section{Discussion}\label{SUMM}

We have shown that locally connected networks of machines
that have primitive learning capabilities
can be used to perform a deterministic, event-based simulation
of quantum computation.
On the other hand it is known that the time evolution of the wave function
of a quantum system can be simulated on a quantum computer~\cite{ZALK98,NIEL00}.
Therefore, it is possible to simulate real-time quantum dynamics
through a deterministic event-based simulation
by constructing appropriate \DLM-networks.
The work presented in this paper suggests that there exist deterministic,
particle-like processes that reproduce quantum mechanical behavior.

Just as any other method for simulating quantum computers~\cite{RAED05},
the \DLM-based simulation approach requires memory resources that increase
exponentially with the number of qubits.
This exponential increase is merely a combinatorial effect
and is not at all related to the quantum nature of the phenomena we want to simulate.
As a matter of fact, it is present in all classical or quantum many-body systems
(including quantum computers).
For instance, if we consider one of the most basic statistical mechanics
models, the Ising model, the number of possible states of this
system also grows exponentially with the number of spins but
there obviously is nothing ``quantum'' about this model.

The computational efficiency of the event-based approach
is lower than the efficiency of algorithms that directly compute
the product of the unitary matrices.
This is hardly a surprise: The former approach simulates quantum behavior
by generating individual events. The latter can only simulate the outcome
of (infinitely) many of such events and provides no information about individual events~\cite{HOME97}.
An analogy may be helpful to understand the conceptual difference between these two approaches.
It is well known that an ensemble of simple, symmetric random walks may
be approximated by a diffusion equation (for vanishing lattice spacing
and time step). Also here we have two options. If we are interested in
individual events, we have no other choice than to simulate the discrete
random walk. However, if we want to study the behavior of many
random walkers, it is computationally much more efficient to solve
the corresponding diffusion problem.

Our event-based approach can be extended to mimic the effects of decoherence.
In quantum theory, decoherence causes the loss of phase coherence~\cite{ZURE03}.
In the \DLMS\ that we describe in Section~\ref{sec4},
a single parameter ($\alpha$) controls the loss of memory, of both
the probability and the phase.
A simple extension would be to control the learning process
of the probability and phase separately, using two control parameters.
We leave this topic for future research.

\section*{Acknowledgement}
We are grateful to Professors M. Imada, S. Miyashita, and M. Suzuki for
many useful comments on the principles of the simulation method
described in this paper.



\raggedright


\begin{thebibliography}{999}
\bibitem{CHIO04}
I. Chiorescu, P. Bertet, K. Semba, Y. Nakamura, C.J.P.M. Harmans, and J.E. Mooij,
Nature {\bf 431}, 159 (2004)

\bibitem{RUGA04}
D. Rugar, R. Budakian, H.J. Mamin, and B.W. Chui,
Nature {\bf 430}, 329 (2004)

\bibitem{ELZE04}
J.M. Elzerman, R. Hanson, L.H. Willems van Beveren, B. Witkamp,
L.M.K. Vandersypen, and L.P. Kouwenhoven,
Nature {\bf 435}, 331 (2004)

\bibitem{XIAO04}
X. Xiao, I. Martin, E. Yablonovitch, and H.W. Jiang,
Nature {\bf 435}, 335 (2004)

\bibitem{FUJI03}
T. Fujisawa,
NTT Technical Review {\bf 1}, 41 (2003)

\bibitem{NIEL00}
M. Nielsen and I. Chuang,
{ Quantum Computation and Quantum Information},
Cambridge University Press, Cambridge (2000)
%
\bibitem{FEYN65}
R.P. Feynman, R.B. Leighton, M. Sands,
{ The Feynman lectures on Physics},
Addison-Wesley, Reading MA (1996), Vol. 3

\bibitem{HOME97}
D. Home, { Conceptual Foundations of Quantum Physics},
Plenum Press, New York (1997)

\bibitem{TONO98}
A. Tonomura,
{ The Quantum World Unveiled by Electron Waves},
World Scientific, Singapore (1998)

\bibitem{BALL03}
L.E. Ballentine,
{ Quantum Mechanics: A Modern Development},
World Scientific, Singapore (2003)

\bibitem{PENR90}
R. Penrose, { The Emperor's New Mind},
Oxford University Press, Oxford (1990)

\bibitem{PerfectExperiments}
In this paper we disregard
limitations of real experiments such as detector efficiency,
imperfection of the source, biprism and the like.

\bibitem{GRAN86}
P. Grangier, R. Roger, and A. Aspect,
Europhys. Lett. {\bf 1}, 173 (1986)
%
\bibitem{KAMP88}
N.G. Van Kampen,
Physica A {\bf 153}, 97 (1988)

\bibitem{KOEN04}
K. De Raedt, H. De Raedt, and K. Michielsen,
http://arxiv.org/abs/quant-ph/0409213

\bibitem{SHOR99}
P.W. Shor,
SIAM Review {\bf 41}, 303 (1999)

\bibitem{HAYK99}
S. Haykin,
{ Neural Networks},
Prentice Hall, New Jersey (1999)

\bibitem{SCHI68}
L.I. Schiff,
{ Quantum Mechanics},
McGraw-Hill, New York (1968)

\bibitem{BAYM74}
G. Baym,
{ Lectures on Quantum Mechanics},
W.A. Benjamin, Reading MA (1974)

\bibitem{QuantumTheory}
We make a distinction between quantum theory and quantum physics.
We use the term {\sl quantum theory} when we refer to
the mathematical formalism, i.e., the postulates
of quantum mechanics (with or without the wave function
collapse postulate)~\cite{BALL03} and the rules (algorithms) to compute the
wave function.
The term {\sl quantum physics} is used for microscopic, experimentally observable phenomena
that do not find an explanation within the mathematical
framework of classical mechanics.

\bibitem{DIVI95a}
D.P. DiVincenzo,
Phys. Rev. A{\bf 51}, 1015 (1995)

\bibitem{BORN64}
M. Born and E. Wolf,
{ Principles of Optics},
Pergamon, Oxford (1964)

\bibitem{RARI97}
J.G. Rarity and P.R. Tapster,
Phil. Trans. R. Soc. Lond. A {\bf 355}, 2267 (1997)

\bibitem{BARE95}
A. Barenco, C.H. Bennett, R. Cleve, D.P. DiVincenzo, N. Margolus,
P. Shor, T. Sleator, J.A. Smolin, and H. Weinfurter,
Phys. Rev. A{\bf 52}, 3457 (1995)

\bibitem{SYPE01b}
L.M.K. Vandersypen,
http://arxiv.org/abs/quant-ph/0205193 (URL last accessed on July 29, 2004)

\bibitem{SYPE01a}
L.M.K. Vandersypen, M. Steffen, G. Breyta, C.S. Yannoni, M.H. Sherwood, and I.L. Chuang,
Nature {\bf 414}, 883 (2001)

\bibitem{EKER96}
A. Ekert and R. Jozsa,
Rev. Mod. Phys. {\bf 68}, 733 (1996)

\bibitem{RAED00}
H. De Raedt, A.H. Hams, K. Michielsen, and K. De Raedt,
Comp. Phys. Comm. {\bf 132}, 1 (2000)

\bibitem{MICH03}
K.F.L. Michielsen and H. De Raedt,
Turk. J. Phys. {\bf 27}, 343 (2003)

\bibitem{RAED05}
H. De Raedt and K. Michielsen,
in {Handbook of Computational and Theoretical Nanotechnology},
Edited by M. Rieth and W. Schommers,
American Scientific Publisher, Los Angeles (2005), in press

\bibitem{QCEDOWNLOAD}
QCE can be downloaded from \url{http:/www.compphys.org/qce.htm}
(URL last accessed on November 23, 2004)

\bibitem{QMvideo}
A large collection of video's of such simulations can be found
at \url{http://www.compphys.org/quantummechanics}
(URL last accessed on November 23, 2004)

%
\bibitem{HANS05}
H. De Raedt, K. De Raedt, and K. Michielsen,
J. Phys. Soc. Jpn. (in press)

\bibitem{MZIdemo}
Interactive programs that performs the event-based simulations
of a beam splitter, one Mach-Zehnder interferometer, and
two chained Mach-Zehnder interferometers can be found at
\url{http://www.compphys.net/dlm}
(URL last accessed on November 23, 2004)

\bibitem{ZALK98}
C. Zalka,
Proc. R. Soc. Lond. A{\bf 454}, 313 (1998)

\bibitem{ZURE03}
W.H. Zurek,
Rev. Mod. Phys. {\bf 75}, 715 (2003)

\end{thebibliography}
\end{document}